\documentclass[a4paper,11pt]{scrartcl}

\usepackage{graphicx}
\usepackage{epsf}
\usepackage[utf8]{inputenc}
\usepackage[T1]{fontenc}
\usepackage[english]{babel}
\usepackage{lmodern}
\usepackage{amsmath,amssymb}
\usepackage{icomma}
\usepackage{bbm}
\usepackage{url}
\usepackage[top=3.75cm,bottom=3.25cm,left=2.25cm,right=3.00cm]{geometry}

\newcommand{\nc}{\newcommand}
\nc{\no}{\nonumber}
\nc{\FS}{F_k^{(S)}}
\nc{\Fb}{F_k^{(b)}}
\nc{\FnS}{F_0^{(S)}}
\nc{\FJ}{F_k^{(J)}}
\nc{\FnJ}{F_0^{(J)}}
\nc{\LS}{L_k^{(S)}}
\nc{\Lb}{L^{(b)}}
\nc{\LJ}{L_k^{(J)}}
\nc{\fS}{f_k^{(S)}}
\nc{\fb}{f_k^{(b)}}
\nc{\fJ}{f_k^{(J)}}
\nc{\TLS}{L^{(S)}}
\nc{\TLJ}{L^{(J)}}
\nc{\nus}{\nu^{(S)}}
\nc{\nuj}{\nu^{(J)}}
\nc{\mns}{m^{(S)}_{0,k}}
\nc{\mes}{m^{(S)}_{1,k}}
\nc{\mej}{m^{(J)}_{1,k}}
\nc{\mzs}{m^{(S)}_{2,k}}
\nc{\mzj}{m^{(J)}_{2,k}}
\nc{\mnsn}{m^{(S)}_{0,0}}
\nc{\mesn}{m^{(S)}_{1,0}}
\nc{\mejn}{m^{(J)}_{1,0}}
\nc{\mzsn}{m^{(S)}_{2,0}}
\nc{\mzjn}{m^{(J)}_{2,0}}
\nc{\erf}{\operatorname{erf}}
\nc{\diag}{\operatorname{diag}}
\nc{\tr}{\operatorname{tr}}

\begin{document}
\setlength{\unitlength}{1mm}
\title{Extreme portfolio loss correlations in credit risk}
\author{Andreas M{\"u}hlbacher\footnote{andreas.muehlbacher@uni-due.de}~ and Thomas Guhr\\{\normalsize Faculty of Physics, University of Duisburg-Essen, Lotharstr. 1, 47048 Duisburg, Germany}}
\date{\today}

\maketitle
\begin{abstract}
The stability of the financial system is associated with systemic risk factors such as the concurrent default of numerous small obligors. Hence it is of utmost importance to study the mutual dependence of losses for different creditors in the case of large, overlapping credit portfolios. We analytically calculate the multivariate joint loss distribution of several credit portfolios on a non-stationary market. To take fluctuating asset correlations into account we use an random matrix approach which preserves, as a much appreciated side effect, analytical tractability and drastically reduces the number of parameters. We show that for two disjoint credit portfolios diversification does not work in a correlated market. Additionally we find large concurrent portfolio losses to be rather likely. We show that significant correlations of the losses emerge not only for large portfolios with thousands of credit contracts but also for small portfolios consisting of a few credit contracts only. Furthermore we include subordination levels, which were established in collateralized debt obligations to protect the more senior tranches from high losses. We analytically corroborate the observation that an extreme loss of the subordinated creditor is likely to also yield a large loss of the senior creditor.
\end{abstract}

\section{Introduction}
The subprime crisis 2007--2009 had a drastic influence on the world economy, due to the almost concurrent default of many small debtors. Most of the credit contracts where bundled into credit portfolios in the form of collateralized debt obligations (CDOs). Realistic estimates for credit risks and the possible losses, particularly of large portfolios are important not only for the creditors, also and maybe even more from a systemic viewpoint. There is a wealth of studies on credit risk, see Refs.~\cite{bielecki2013,bluhm2016,crouhy2000,lando2009,mcneil2015} and references therein.

In a credit portfolio it is of utmost importance to consider the correlations of the asset values. It has been shown that in the presence of even little correlations the concept of diversification is deeply flawed, see Refs.~\cite{schaefer2007,schmitt2014EPL,schmitt2015CR}. Hence it is not possible to lower the tail risk significantly by enlarging the number of credit contracts in a credit portfolio. In general, diversification is not always fruitful \cite{ibragimov2007,wagner2010}.

To obtain a comprehensive understanding of systemic credit risk, it is important to study and model the mutual dependence of losses of different portfolios. Here we are interested in the joint probability distribution that contains all the information on the individual loss distributions as well as their dependence structure. We apply the Merton model \cite{merton1974pricing,muennix2014} to several credit portfolios simultaneously. Additionally we take fluctuating asset correlations into account. These emerge because of the intrinsic non-stationarity of financial markets which leads to a change of the correlation and covariance matrix in time \cite{munnix2012,schmitt2013EPL,song2011}. To describe this non-stationarity we use an ensemble approach which was recently introduced in Ref.~\cite{chetalova2015}. It results in a multivariate asset return distribution averaged over the fluctuating correlation matrices. The validity of this approach has been confirmed by empirical data analysis \cite{schmitt2015CR,schmitt2013EPL}. The ensemble approach leads to a drastic reduction of the number of parameters describing the distribution. Remarkably, only two parameters, the average correlation level of the asset values and the strength of the fluctuations are sufficient. From the asset return distribution we analytically derive a joint probability distribution of credit portfolio losses. In addition we derive a limiting distribution for infinitely large credit portfolios. We analyze in detail two non-overlapping credit portfolios that operate on the same market. Moreover we include subordination levels \cite{an2015subordination,blackcox1976,gorton1990market}. At maturity time the senior creditor is paid out first and the junior subordinated creditor is only paid out if the senior creditor regained the full promised payment. This is related to CDO tranches and gives further information on to multivariate credit risk \cite{duffie2001,longstaff2008}.

Furthermore, we consider a single credit portfolio that operates on several markets which are on average uncorrelated. We are able to derive a limiting distribution for an infinitely large credit portfolio. Here, the tail risk is lower than in the case of one market with homogeneous correlation structure, but still diversification is limited.

The paper is organized as follows. In section \ref{Sec model} we introduce the Merton model and derive the portfolio loss distributions for different debt structures. In section \ref{Sec results} we present our results for empirical estimated parameters. We conclude our observations in section \ref{Sec conclusion}.

\section{Model}\label{Sec model}
\begin{figure}[ht]
 \centering
 \includegraphics[width=0.5\textwidth]{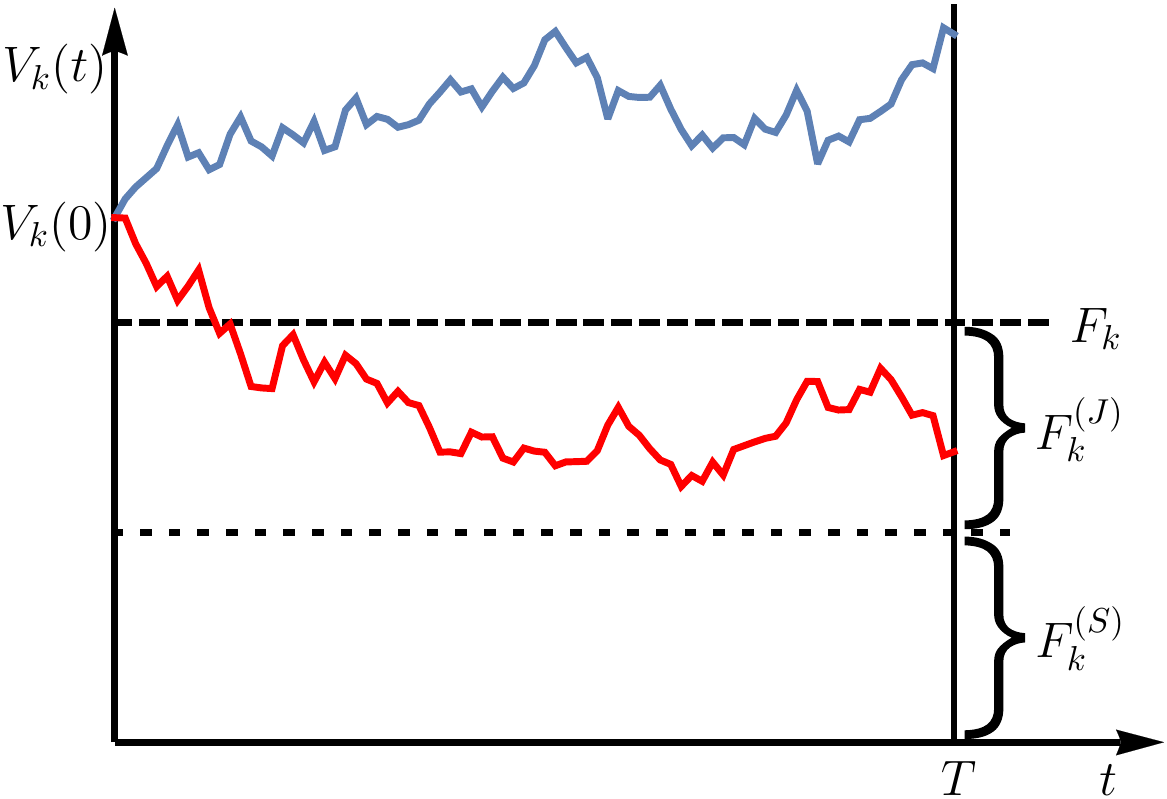}
  \caption{Schematic visualization of the Merton model. A default occurs if the asset value at maturity $V_K(T)$ drops below the face value $F_k$. In the red sketched scenario a default occurs only to the junior subordinated creditor while the senior creditor obtains no loss.}
 \label{fig MertonM}
\end{figure}

We extend the Merton model to a multivariate scenario with two creditors and $K$ correlated obligors with asset values or economic states $V_k(t)$, $k=1,\dots,K$ at time $t$. Each obligor may hold a credit contract from each creditor. In the Merton model the asset values $V_k(t)$ are estimated by the stock prices of the corresponding obligors. So we assume that all $K$ obligors are companies which can be traded on a stock market. We claim that the asset values follow a geometric Brownian motion. Further, we assume subordinated debt where at maturity time $T$ the senior creditor is paid out first and the junior subordinated creditor is only paid out if the senior creditor regained the full promised payment. Suppose each obligor has to pay back the face value $F_k$ at maturity time $T$. We consider large time scales such as one year or one month. The face value of each obligor is composed of the face value of the senior creditor $\FS$ and the face value of the junior subordinated creditor $\FJ$, that is $F_k=\FS+\FJ$. A default occurs if the asset value drops below the face value i.e. $V_k(T)<F_k$ for at least one obligor. The severity of the loss depends on the value of the obligors $V(T)$ at time maturity. For $F_k>V_k(T)>\FS$ the default is completely defrayed by the junior subordinated creditor meaning that the senior creditor does not incur any loss. Only if $V_k(T)<\FS$ the senior creditor will incur a loss while the junior subordinated creditor will sustain a total loss. A visualization of the underlying process for a single asset is shown in Fig.~\ref{fig MertonM}.
The normalized loss $L_k^{(S)}$ that a senior creditor and the normalized loss $\LJ$ that a junior subordinated creditor is suffering can be expressed as
\begin{align}
\LS =&\left(1-\frac{V_k(T)}{\FS}\right)\Theta\left(\FS-V_k(T)\right)\label{eq lossesdefSen}\\
\LJ =&\left(1-\frac{V_k(T)-\FS}{\FJ}\Theta\left(V_k(T)-\FS\right)\right)\Theta\left(F_k-V_k(T)\right)\;,
\label{eq lossesdefJun}
\end{align}
respectively. The Heaviside step functions $\Theta(x)$ ensure that the losses are strictly positive. We introduce the fractional face values $f_k^{(S)}$ and $\fJ$ for the senior and junior subordinated creditors
\begin{align}
\fS=\frac{\FS}{\sum^{K}_{l=1}{F_l^{(S)}}} \hspace{1cm}\text{and}\hspace{1cm}\fJ=\frac{\FJ}{\sum^{K}_{l=1}{F_l^{(J)}}}\;,
\end{align}
respectively. This enables us to define the normalized portfolio losses $L^{(S)}$ and $L^{(J)}$ for the senior and junior subordinated creditors as weighted sums
\begin{align}
\TLS=\sum^{K}_{k=1}{\fS \LS}\hspace{1cm}\text{and}\hspace{1cm}\TLJ=\sum^{K}_{k=1}{\fJ \LJ}\;,
\end{align}
respectively. Our aim is to derive the bivariate distribution $p(\TLS,\TLJ)$ of the portfolio losses. This can be done by integrating over all portfolio values and filtering those that lead to a given bivariate total loss $(\TLS,\TLJ)$
\begin{align}
p(\TLS,\TLJ)=\int{d[V]g(V|\Sigma)\,\delta\!\left(\TLS-\sum^{K}_{k=1}{\fS \LS}\right)\!\delta\!\left(\TLJ-\sum^{K}_{k=1}{\fJ \LJ}\right)}\;,
\label{eqlossdistgeneral}
\end{align}
where $g(V|\Sigma)$ is the multivariate distribution of the correlated asset values of the obligors at maturity, $\Sigma$ is the covariance matrix of the asset values, which is in our model well estimated by that of the stock prices. $\delta(x)$ is the Dirac delta function and $V=(V_1(T),\dots,V_K(T))$ is the $K$ component vector of the asset values. The measure $d[V]$ is the product of all differentials and the integration domain ranges from zero to infinity for every integral. Using the Fourier representation of the $\delta$ function as well as Eqs.~\eqref{eq lossesdefSen} and \eqref{eq lossesdefJun}, we find
\begin{align}
p(\TLS,\TLJ)&=\frac{1}{(2\pi)^2}\int\limits^{\infty}_{-\infty}{d\nus e^{-i\nus \TLS}}\int\limits^{\infty}_{-\infty}{d\nuj e^{-i\nuj \TLJ}}\no \\
&\quad\times\prod^{K}_{k=1}\left[\int\limits^{\FS}_{0}{dV_k}\exp\left(i\nus\fS\left(1-\frac{V_k}{\FS}\right)+i\nuj\fJ\right)\right.\no\\
&\quad+\left.\int\limits^{F_k}_{\FS}{dV_k\exp\left(i\nuj\fJ\left(1-\frac{V_k-\FS}{\FJ}\right)\right)+\int\limits^{\infty}_{F_k}}{dV_k}\right]g(V|\Sigma)\;,
\label{eq loss3ints}
\end{align}
where we split the $V_k$ integrals in three parts. We will use this expression later on, but we first need to specify the multivariate distribution of the correlated asset values $g(V|\Sigma)$.

Our goal is to calculate joint loss distributions which take the non-stationarity of the covariances into account,
\begin{align}
\langle p\rangle(\TLS,\TLJ)=\int{d[V]\langle g\rangle(V|\Sigma)\,\delta\!\left(\TLS-\sum^{K}_{k=1}{\fS \LS}\right)\!\delta\!\left(\TLJ-\sum^{K}_{k=1}{\fJ \LJ}\right)}
\label{eqlossdistgeneralAverage}
\end{align}
and according for Eq.~\eqref{eq loss3ints}. We will argue that this is achieved by properly averaging the multivariate distribution $g(V|\Sigma)$, resulting in $\langle g\rangle(V|\Sigma)$.

\subsection{Average distribution}
Following Refs.~\cite{schmitt2015CR,muennix2014,schmitt2013EPL}, we use a random matrix concept to capture the non-stationarity of the correlations between the asset values $V_k$. The covariances of the returns
\begin{align}
r_k(t)=\frac{V_k(t+\Delta t)-V_k(t)}{V_k(t)}\;,
\end{align}
with the return interval $\Delta t$ are ordered in the $K\times K$ covariance matrix $\Sigma$. It can be expressed as $\Sigma=\sigma C \sigma$ with the correlation matrix $C$ and the diagonal matrix $\sigma=\diag\left(\sigma_1,\dots,\sigma_K\right)$ containing the volatilities of the different return time series $r_k$. As the covariance matrices differ significantly for different times, we obtain an existing ensemble of different covariance matrices. In Refs.~\cite{schmitt2015CR,muennix2014,schmitt2013EPL} it was demonstrated that this data ensemble can be very efficiently modeled by random covariance or correlation matrices distributed according to Wishart \cite{wishart1928generalised}
\begin{align}
 w(W|\Sigma,N)=\frac{\sqrt{N}^{KN}}{\sqrt{\det(2\pi\Sigma)}^N}\exp\left(-\frac{N}{2}\tr W^\dagger\Sigma^{-1}W\right)\;.
\label{eq wishart}
\end{align}
This distribution defines an ensemble of random covariance matrices $WW^\dagger$. They fluctuate around the average covariance matrix $\Sigma$ which is empirically evaluated over the whole time interval. The symbol $\dagger$ denotes the transpose of a vector or matrix. The model matrices $W$ have dimension $K\times N$. Here the free parameter $N$ corresponds to the length of the model time series, it has to be determined from the data. $N$ controls the strength of the fluctuations around the average covariance matrix $\Sigma$. The smaller $N$, the larger are the fluctuations. The ensemble average leads to the following general result for the average return distribution in the presence of fluctuating covariance matrices,
\begin{align}
\left\langle g\right\rangle\left(r|\Sigma,N\right)=\frac{\sqrt{N}^K}{\sqrt{2}^{N-2}\Gamma(N/2)\sqrt{\det(2\pi\Sigma)}}\frac{\mathcal{K}_{(K-N)/2}\left(\sqrt{Nr^\dagger\Sigma^{-1}r}\right)}{\sqrt{Nr^\dagger\Sigma^{-1}r}^{(K-N)/2}}\:,
\label{eq dist rets}
\end{align}
where $\mathcal{K}_{(K-N)/2}$ is the Bessel function of second kind and of order $(K-N)/2$ \cite{schmitt2013EPL}. $\Sigma$ in Eq.~\eqref{eq dist rets} is the average over the whole data interval considered. In this notation and all further notations we omit the time dependences of $r$ and $V$. Since we assume all credit contracts to have the form of zero coupon bonds, we consider our return intervals to have the same length as the maturity time, i.e. $\Delta t=T$.

In Refs.~\cite{schmitt2015CR,schmitt2013EPL} it has been shown that an effective average correlation matrix of the form
\begin{align}
C=(1-c)\mathbbm{1}_K+ce_Ke_K^\dagger=\begin{bmatrix}
	1 & c & c & \dots\\
	c & 1 & c & \dots\\
	c & c & 1 & \dots\\
	\vdots & \vdots & \vdots & \ddots
\end{bmatrix}
\label{eq avcorrm}
\end{align}
with $\mathbbm{1}_K$ being the $K\times K$ unit matrix and $e_K$ being a $K$ component vector containing ones, yields a good description of empirical data in the present setting. If we studied non-averaged quantities depending on a specific correlation structure, this approach would be much less likely to give satisfactory results. The choice has two major advantages. First we achieve analytical tractability what can be seen later on in section \ref{Sec ALD}, second we can describe the complexity of a correlated market with only two parameters. The first parameter $c$ is an effective average correlation level and the second parameter $N$ describes the strength of the fluctuations around this average. Both parameters have to be estimated from empirical data.

Due to the fact that we need the asset values $V_k(T)$ in our loss distribution \eqref{eqlossdistgeneral} while covariances are measured with returns, we have to perform a change of variables using Ito's lemma \cite{ito}
\begin{align}
r_k\rightarrow \ln\frac{V_k(T)}{V_{k0}}-\left(\mu_k-\frac{\rho^2_k}{2}\right)T\;,
\end{align}
where $V_{k0}=V_k(0)$ is the initial asset value. This is a geometric Brownian motion with drift $\mu_k$ and standard deviation $\rho_k$ with $\sigma_k=\rho_k\sqrt{T}$. The expression \eqref{eq dist rets} can now be rewritten using Fourier integrals. After employing and adjusting the steps in Ref.~\cite{schmitt2014EPL}, we arrive at the double integral
\begin{align}
\left\langle g\right\rangle\left(V|c,N\right)&=\frac{1}{2^{N/2}\Gamma(N/2)}\int\limits^{\infty}_{0}{dz z^{N/2-1}e^{-z/2}\sqrt{\frac{N}{2\pi z}}\sqrt{\frac{N}{2\pi z(1-c)T}}^K}\int\limits^{\infty}_{-\infty}{du\exp\left(-\frac{N}{2z}u^2\right)} \no \\
&\quad\times  \prod^{K}_{k=1}{\frac{1}{V_k\rho_k}\exp\left[-\frac{N}{2z(1-c)T\rho^{2}_{k}}\left(\ln\frac{V_k}{V_{k0}}-\left(\mu_k-\frac{\rho^{2}_{k}}{2}\right)T+\sqrt{cT}u\rho_k\right)^2\right]}\;.
\label{eq assetdist}
\end{align}
The random matrix model of non-stationarity together with the effective average correlation matrix results in an expression for the joint multivariate distribution of the asset values in terms of a bivariate average of the product of geometric Brownian motions over a $\chi^2$ distribution in $z$ and a Gaussian in $u$. We do not perform the $u$ integration yet, because we will factorise the $V_k$ integrals when computing the loss distribution \eqref{eqlossdistgeneral} later on.

\subsubsection{Several on average uncorrelated markets}
To be more realistic we consider not just one market but several markets which are on average uncorrelated. This is an extension of unpublished work by T.~Nitschke \cite{nitschke}. We define the number of uncorrelated markets to be $\beta$. In this case the correlation matrix $C=\diag(C_1,\dotsc,C_\beta)$ is block diagonal where $C_l=(1-c_l)\mathbbm{1}_{K_l}+c_le_{K_l}e_{K_l}^\dagger$ are matrices themselves with dimensions $K_l\times K_l$ for $l\in\{1,\dotsc,\beta\}$. The correlation matrix $C$ has dimension $K\times K$ and therefore $\sum^{\beta}_{l=1}K_l=K$ holds. This block structure in not reflected in the random correlation matrices fluctuating about $C$, see Eq.~\eqref{eq wishart}. Hence, there are correlations between the blocks, only their average is zero. The correlation structure allows us to study the impact when going from one market to several markets. Within one market we again have an on average effectively correlation structure and across the markets we have an average correlation of zero. Importantly, this only means the absence of correlations on average. The correlations in our model and in reality fluctuate, implying that in any short instant of time, correlations can be present whose strength is governed by the parameter $N$. Furthermore each market has its own volatility matrix $\sigma_l=\diag(\sigma_{l1},\dotsc,\sigma_{lK_l})$ and drift vector $\mu_l=(\mu_{l1},\dotsc,\mu_{lK_l})^\dagger$ for $l\in\{1,\dotsc,\beta\}$. We properly extend the calculations in Ref.~\cite{schmitt2014EPL} with the difference that we have to apply $l$ Fourier integrals, yielding
\begin{align}
\left\langle g\right\rangle\left(V|c,N\right)&=\frac{1}{2^{N/2}\Gamma(N/2)}\int\limits^{\infty}_{0}{dz z^{N/2-1}e^{-z/2}\sqrt{\frac{N}{2\pi z}}^\beta\sqrt{\frac{N}{2\pi zT}}^K\left(\prod^{\beta}_{l=1}\prod^{K_l}_{k=1}\frac{1}{V_{lk}\rho_{lk}}\right)} \no \\
&\quad\times  \prod^{\beta}_{l=1}\frac{1}{\sqrt{1-c_l}^{K_l}}\int\limits^{\infty}_{-\infty}{du_l\exp\left(-\frac{N}{2z}u^2_l\right)} \no \\
&\quad\times  \exp\left[-\frac{N}{2zT}\sum_k\frac{\left(\ln\frac{V_{lk}}{V_{lk0}}-\left(\mu_{lk}-\frac{\rho^{2}_{lk}}{2}\right)T+\sqrt{c_lT}u_l\rho_{lk}\right)^2}{(1-c_l)\rho^2_{lk}}\right]\;.
\label{eq assetdistseveralmarkets}
\end{align}
This multiple integral depends on the number of markets $\beta$. The index $l$ indicates each market, the index $k$ indicates the asset in a specified market $l$. In general, the index pair $(l,k)$ denotes the $k$th asset on the $l$th market.

\subsection{Average loss distribution}\label{Sec ALD}
We work out the average loss distribution \eqref{eqlossdistgeneralAverage} using the above results for the average distribution $\left\langle g\right\rangle(V|c,N)$. After inserting Eq.~\eqref{eq assetdist} into Eq.~\eqref{eq loss3ints} we obtain
\begin{align}
\left\langle p\right\rangle(\TLS,\TLJ)&=\frac{1}{(2\pi)^22^{N/2}\Gamma(N/2)}\int\limits^{\infty}_{0}{dz z^{N/2-1}e^{-z/2}\sqrt{\frac{N}{2\pi z}}\int\limits^{\infty}_{-\infty}{du\exp\left(-\frac{N}{2z}u^2\right)}}\no\\
&\quad\times \int\limits^{\infty}_{-\infty}{d\nus e^{-i\nus \TLS}\int\limits^{\infty}_{-\infty}{d\nuj e^{-i\nuj \TLJ}\mathcal{I}\left(\nus,\nuj,z,u\right)}}\;,
\end{align}
with the term
\begin{align}
\mathcal{I}\left(\nus,\nuj,z,u\right)=\prod^{K}_{k=1}{\left\{1+\sum^{\infty}_{j=1}{\frac{i^j}{j!}m^{(SD)}_{j,k}\left(\nus,\nuj,z,u\right)}+\sum^{\infty}_{j=1}{\frac{\left(i\nuj\fJ\right)^j}{j!}m^{(J)}_{j,k}\left(z,u\right)}\right\}}
\end{align}
and
\begin{align}
m^{(SD)}_{a,k}\left(\nus,\nuj,z,u\right)=\sum^{a}_{j=0}\binom{a}{j}\left(\nus\fS\right)^j\left(\nuj\fJ\right)^{a-j}m^{(S)}_{j,k}(z,u)
\end{align}
and the moments
\begin{align}
m^{(S)}_{j,k}(z,u)&=\int\limits^{\hat{F}^{(S)}_k}_{-\infty}{d\hat{V}_k\left(1-\frac{V_{k0}}{\FS}\exp\left(\sqrt{z}\hat{V}_k+\left(\mu_k-\frac{\rho^2_k}{2}\right)T\right)\right)^j}\no\\
&\quad\times \sqrt{\frac{N}{2\pi(1-c)T\rho^2_k}}\exp\left[\frac{N}{2(1-c)T\rho^2_k}\left(\hat{V}_k+\sqrt{cT}u\rho_k\right)^2\right]\label{eq moment}\\
m^{(J)}_{j,k}(z,u)&=\int\limits^{\hat{F}_{k}}_{\hat{F}^{(S)}_k}{d\hat{V}_k\left(1+\frac{\FS}{\FJ}-\frac{V_{k0}}{\FJ}\exp\left(\sqrt{z}\hat{V}_k+\left(\mu_k-\frac{\rho^2_k}{2}\right)T\right)\right)^j}\no\\
&\quad\times \sqrt{\frac{N}{2\pi(1-c)T\rho^2_k}}\exp\left[\frac{N}{2(1-c)T\rho^2_k}\left(\hat{V}_k+\sqrt{cT}u\rho_k\right)^2\right]\;,
\end{align}
where we use the change of variables $\hat{V}_k=(\ln V_k/V_{k0}-(\mu_k-\rho^2_k/2)T)/\sqrt{z}$ with proper adjustment of the integration bounds $\hat{F}_{k}$ and $\hat{F}^{(S)}_{k}$. The moments $m^{(S)}_{j,k}(z,u)$ and $m^{(J)}_{j,k}(z,u)$ are given in appendix \ref{app: moments} for $j=0,1,2$. The term $m^{(SD)}_{j,k}\left(\nus,\nuj,z,u\right)$ formally corresponds to those events that lead to a loss large enough that the senior creditor is affected. We use a binomial sum for the decoupling of the $\nus$ and $\nuj$ integrals later on.

Now we assume large portfolios where all face values are of the same order, to carry out an approximation to the second order in $\fS$ and $\fJ$ by performing steps generalizing the one in Ref.~\cite{schmitt2014EPL}. This is justified when we consider all face values are of the same order, so all fractational face values are of order $1/K$. We finally arrive at
\begin{align}
\left\langle p\right\rangle(\TLS,\TLJ)&=\frac{1}{2^{N/2}\Gamma(N/2)}\int\limits^{\infty}_{0}{dz z^{N/2-1}e^{-z/2}\sqrt{\frac{N}{2\pi }}\int\limits^{\infty}_{-\infty}{du\exp\left(-\frac{N}{2}u^2\right)}}\no\\
&\quad\times \frac{1}{\sqrt{2\pi M^{(S)}_2(z,u)}}\exp\left(-\frac{\left(\TLS-M^{(S)}_1(z,u)\right)^2}{2M^{(S)}_2(z,u)}\right)\no\\
&\quad\times \frac{1}{\sqrt{2\pi M_2(z,u)}}\exp\left(-\frac{\left(\TLJ-M_1(\TLS,z,u)\right)^2}{2M_2(z,u)}\right)
\label{eq lossdistsubfin}
\end{align}
for the average distribution with
\begin{align}
M_1(\TLS,z,u)&=M^{(J)}_1(z,u)+\sum\limits^{K}_{k=1}\fJ\fS N^{(S)}_k(z,u)\frac{\TLS-M^{(S)}_1(z,u)}{M^{(S)}_2(z,u)}\\
M_2(z,u)&=M^{(J)}_2(z,u)-\frac{1}{M^{(S)}_2(z,u)}\left(\sum\limits^{K}_{k=1}\fJ \fS N^{(S)}_k(z,u)\right)^2\label{eq M2}\\
M^{(S)}_1(z,u)&=\sum\limits^{K}_{k=1}{\fS \mes}\\
M^{(S)}_2(z,u)&=\sum\limits^{K}_{k=1}{\fS}^2\left(\mzs-{\mes}^2\right)\label{eq M2S}\\
M^{(J)}_1(z,u)&=\sum\limits^{K}_{k=1}{\fJ\left(\mns+\mej\right)}\\
M^{(J)}_2(z,u)&=\sum\limits^{K}_{k=1}{{\fJ}^2\left(\mns+\mzj-{\mns}^2-{\mej}^2-2\mns\mej\right)}\\
N^{(S)}_k(z,u)&=\mes\left(1-\mns-\mej\right)\;.
\end{align}
Thus, we expressed the average loss distribution as double average of Gaussians with mean values $M_1(\TLS,z,u)$ and $M^{(S)}_1(z,u)$ and variances $M_2(z,u)$ and $M^{(S)}_2(z,u)$ that non-trivially depend on the integration variables. To keep the notation transparent, we dropped the arguments of the functions $m^{(S)}_{j,k}(z,u)$ and $m^{(J)}_{j,k}(z,u)$. Due to the complexity of the last two expressions in Eq.~\eqref{eq lossdistsubfin} the $z$ and $u$ integrals have to be evaluated numerically. We notice that the normalization of the average distribution is for $\LS,\LJ\in[0,1]$ only valid up to the order of our approximation. Later on we will concentrate on the contributions of no default.

\subsection{Homogeneous portfolio}
Apart from the large $K$ approximation, all results above are valid in general and apply to all portfolios for which the individual fractational face values are of order $1/K$. To further evaluate our results and to obtain a visualization, it is instructive to consider homogeneous portfolios, in which the senior and junior face values are equal,
\begin{align}
 \FS=\FnS\hspace{1cm} \text{and}\hspace{1cm} \FJ=\FnJ
\end{align}
such that
\begin{align}
\fS=\fJ=\frac{1}{K}\;.
\end{align}
Furthermore, we assume that the stochastic processes have the same initial values, drift and standard deviations,
\begin{align}
 V_{k0}=V_0\;,\hspace{1cm}\mu_k=\mu_0\;,\hspace{1cm}\rho_k=\rho_0\;.
\end{align}
Of course, this does not mean that the realized stochastic processes are the same. By dropping the dependence of $k$, the moments $m^{(S)}_{a,k}(z,u)=m^{(S)}_{a,0}(z,u)$ and $m^{(J)}_{j,k}(z,u)=m^{(J)}_{j,0}(z,u)$ and thus the average distribution $\left\langle p\right\rangle(\TLS,\TLJ)$ can be computed much faster.

\subsection{Distribution of the loss given default}
Only the full dynamics of our model without any approximations gives us information on the contribution of the non-analytic part of the average loss distribution. In particular, absence of losses is reflected in non-analytic $\delta$ functions at zero. To examine this we start from the averaged version of Eq.~\eqref{eq loss3ints} by inserting the distribution of asset values for a homogeneous portfolio with an effective average correlation matrix
\begin{align}
\left\langle g\right\rangle_{\text{h}}\left(V|c,N\right)&=\frac{1}{2^{N/2}\Gamma(N/2)}\int\limits^{\infty}_{0}{dz z^{N/2-1}e^{-z/2}\sqrt{\frac{N}{2\pi z}} \int\limits^{\infty}_{-\infty}{du\exp\left(-\frac{N}{2z}u^2\right)}}\no\\
&\times\left(\!\sqrt{\!\frac{N}{2\pi z(1\!-\!c)T}}\frac{1}{V\rho_0} \exp\!\left[-\frac{N}{2z(1\!-\!c)T\rho^{2}_{0}}\left(\!\ln\!\frac{V}{V_{0}}-\left(\!\mu_0-\frac{\rho^{2}_{0}}{2}\right)T+\sqrt{cT}u\rho_0\!\right)^{\!\!2}\right]\right)^{\!\!\!K}\no \\
&=\int\limits^{\infty}_{0}{dz \int\limits^{\infty}_{-\infty}{du f(z,u) \tilde{\omega}^K(V,z,u)}},
\end{align}
with
\begin{align}
 f(z,u)&=\frac{1}{2^{N/2}\Gamma(N/2)}z^{N/2-1}e^{-z/2}\sqrt{\frac{N}{2\pi z}}\exp\left(-\frac{N}{2z}u^2\right)\\
 \tilde{\omega}(V,z,u)&=\sqrt{\!\frac{N}{2\pi z(1\!-\!c)T}}\frac{1}{V\rho_0} \exp\!\left[-\frac{N}{2z(1\!-\!c)T\rho^{2}_{0}}\left(\!\ln\!\frac{V}{V_{0}}-\left(\!\mu_0-\frac{\rho^{2}_{0}}{2}\right)T+\sqrt{cT}u\rho_0\!\right)^{\!\!2}\right].
\end{align}
Due to the homogeneity, the product in Eq.~\eqref{eq loss3ints} also becomes a $K$-th power, to which we apply the multinomial theorem. We thus arrive at
\begin{align}
\left\langle p\right\rangle(\TLS,\TLJ)&=\int\limits^{\infty}_{0}{dz \int\limits^{\infty}_{-\infty}{du f(z,u)\frac{1}{(2\pi)^2}\int\limits^{\infty}_{-\infty}{d\nus e^{-i\nus \TLS}\int\limits^{\infty}_{-\infty}{d\nuj e^{-i\nuj \TLJ}}}}}\no\\
&\quad\times\!\!\!\!\!\!\!\!\! \sum\limits_{k_1+k_2+k_3=K}{\binom{K}{k_1,k_2,k_3}\!\!\left(e^{i\nuj/K}\!\!\int\limits^{\,\FnS}_{0}{dV}\exp\left(\frac{i\nus}{K}\left(1-\frac{V}{\FnS}\right)\right)\tilde{\omega}(V,z,u)\right)^{\!\!\!k_1}}\no\\
&\quad\times\left(\int\limits^{F_0}_{\FnS}{dV\exp\left(\frac{i\nuj}{K}\left(1-\frac{V-\FnS}{\FnJ}\right)\right)\tilde{\omega}(V,z,u)}\right)^{\!\!\!k_2}\!\!\!\!\left(\int\limits^{\infty}_{F_0}{dV\tilde{\omega}(V,z,u)}\right)^{\!\!\!k_3}
\label{eq exdistrmultinorm}
\end{align}
with the multinomial coefficient
\begin{align}
\binom{K}{k_1,k_2,k_3}=\frac{K!}{k_1!k_2!k_3!}\;.
\end{align}
From Eq.~\eqref{eq exdistrmultinorm} we see that $\delta$ functions only appear under the condition $k_1\cdot k_2=0$. For $k_1=k_2=0$ we have no default at all, the only contribution to the distribution stems from the last integral in Eq.~\eqref{eq exdistrmultinorm} leading to a $\delta$ peak $\delta(\LS)\delta(\LJ)$ at the origin. This $\delta$ peak is associated with the absence of default neither on the junior nor on the senior level. The probability therefore is
\begin{align}
P^{(\text{ND})}&=\frac{1}{2^{N/2}\Gamma(N/2)}\int\limits^{\infty}_{0}{dz z^{N/2-1}e^{-z/2}\sqrt{\frac{N}{2\pi z}} \int\limits^{\infty}_{-\infty}{du\exp\left(-\frac{N}{2z}u^2\right)}}\no\\
&\quad\times\left(\frac{1}{2}-\frac{1}{2}\erf\left[\sqrt{\frac{N}{2z(1-c)T\rho^2_0}}\left(\ln\frac{F_0}{V_0}-\left(\mu_0-\frac{\rho_0^2}{2}\right)T+\sqrt{cT}u\rho_0\right)\right]\right)^K\;.
\label{eq survivalprobab}
\end{align}
It obviously decreases with increasing $K$.

For $k_1=0, k_2\neq0$ we find the contribution of the events that lead to a total junior default but not to a senior default. In this case we have a single $\delta$ function $\delta(\LS)$ which represents a moderate loss such that the senior subordinated creditor will not sustain a loss. The special case $k_1\neq0, k_2=0$ leads to a sum of $\delta$ functions $\delta(\LJ-k_1/K)$ where $k_1$ runs from $1$ to $K$. This is due to the sum in Eq.~\eqref{eq exdistrmultinorm}. These $\delta$ functions belong to the events where there is either no default at all or $k_1$ severe defaults such that for $k_1=1,\dotsc,K$ obligors the junior subordinated creditor has a complete failure i.e. $\LJ=1$ and the senior subordinated creditor may sustain a loss i.e. $\LS\geq0$. All these $\delta$ functions are not unmated but weighted with some integral prefactors to preserve the normalization of the distribution $\left\langle p\right\rangle(\TLS,\TLJ)$. Furthermore the $\delta$ functions disappear when we only consider the loss given default, which in our model means $\TLJ>0$ and also $\TLS>0$. The non-analytic parts cannot be obtained in the second order approximation we used to derive the average loss distribution \eqref{eq lossdistsubfin}.

\subsection{Infinitely large portfolios}
We now consider the case $K\to\infty$ for the homogeneous portfolio to analyze whether diversification works or not in the discussed multivariate scenarios. It has been shown that diversification does not work in a correlated univariate model with only one bank \cite{schmitt2014EPL}.

The homogeneous versions of Eqs.~\eqref{eq M2} and \eqref{eq M2S}
\begin{align}
M_2(z,u)&=\dfrac{1}{K}\left(\mnsn+\mzjn-{\mnsn}^2-{\mejn}^2-2\mnsn\mejn-\frac{{\mesn}^2\left(1-\mnsn-\mejn\right)^2}{\mzsn-{\mesn}^2}\right)\\
M^{(S)}_2(z,u)&=\dfrac{1}{K}\left(\mzsn-{\mesn}^2\right)
\end{align}
imply that $M_2(z,u)\to0$ as well as $M^{(S)}_2(z,u)\to0$ for $K\to\infty$. This means that both Gaussians
\begin{align}
\frac{1}{\sqrt{2\pi M^{(S)}_2\!(z,u)}}\exp\!\left(\!-\frac{\left(\TLS\!-\!M^{(S)}_1\!(z,u)\right)^{\!2}}{M^{(S)}_2(z,u)}\right)\hspace{0.1cm} \text{and}\hspace{0.1cm}\frac{1}{\sqrt{2\pi M_2(z,u)}}\exp\!\left(\!-\frac{\left(\TLJ\!-\!M_1(\TLS,z,u)\right)^{\!2}}{M_2(z,u)}\right)\no
\end{align}
in Eq.~\eqref{eq lossdistsubfin} become $\delta$ functions. Thus we arrive at
\begin{align}
\left\langle p\right\rangle(\TLS,\TLJ)&=\frac{1}{2^{N/2}\Gamma(N/2)}\int\limits^{\infty}_{0}{dz z^{N/2-1}e^{-z/2}\sqrt{\frac{N}{2\pi }}\int\limits^{\infty}_{-\infty}{du\exp\left(-\frac{N}{2}u^2\right)}}\no\\
&\quad\times\delta\left(\TLS-M^{(S)}_1(z,u)\right)\delta\left(\TLJ-M_1(\TLS,z,u)\right)\no\\
&=\frac{1}{2^{N/2}\Gamma(N/2)}\int\limits^{\infty}_{0}{dz z^{N/2-1}e^{-z/2}\sqrt{\frac{N}{2\pi }}\int\limits^{\infty}_{-\infty}{du\exp\left(-\frac{N}{2}u^2\right)}}\no\\
&\quad\times\delta\left(\TLS-\mesn\right)\delta\left(\TLJ-\mnsn-\mejn\right)\;.
\end{align}
To make this equation numerically manageable we use the identity
\begin{align}
\delta\left(f(u)\right)=\sum\limits_i\frac{\delta(u-u_i)}{|f'(u_i)|}\;,
\label{eq deltatrick}
\end{align}
where $u_i$ are the roots of the function $f(u)$, with $f^\prime(u_i)\neq0$. Using this identity three times allows us to solve the remaining two integrals and we finally obtain the limiting loss distribution
\begin{align}
\left\langle p\right\rangle(\TLS,\TLJ)&=\frac{1}{2^{N/2}\Gamma(N/2)}\sqrt{\frac{N}{2\pi}}z^{N/2-1}_{0}\exp\left(-\frac{z_0}{2}\right)\exp\left(-\frac{N}{2}{u^{(S)}}^2(\TLS,z_0)\right)\no\\
&\quad\times\frac{1}{\left|\frac{\partial}{\partial u}\left.\mesn(z_0,u)\right|_{u=u^{(S)}(\TLS,z_0)}\right|\cdot\left|\frac{\partial}{\partial u}\left[\mnsn(z_0,u)+\mejn(z_0,u)\right]_{u=u^{(S)}(\TLS,z_0)}\right|}\no\\
&\quad\times\frac{1}{\left|\frac{\partial}{\partial z}\left[u^{(S)}(\TLS,z)-u^{(J)}(\TLJ,z)\right]_{z=z_0}\right|}\;.
\label{eq limfin}
\end{align}
Here the implicit functions 
\begin{align}
u^{(S)}&=u^{(S)}(\TLS,z) \hspace{1cm}&\text{with}\hspace{1cm}&0=\TLS-\mesn(z,u^{(S)})\label{eq usimpl}\\
u^{(J)}&=u^{(J)}(\TLJ,z) \hspace{1cm}&\text{with}\hspace{1cm}&0=\TLJ-\mnsn(z,u^{(J)})-\mejn(z,u^{(J)})\label{eq ujimpl}\\
z_0&=z_0(\TLS,\TLJ)\hspace{1cm}&\text{with}\hspace{1cm}&u^{(S)}(\TLS,z_0)=u^{(J)}(\TLJ,z_0)\;,
\end{align}
are unique and have to be calculated numerically. The dependence on $\TLS$ and $\TLJ$ is now implicit in the functions $u^{(S)},u^{(J)}$ and $z_0$. The very last derivatives in Eq.~\eqref{eq limfin} can be done by using the implicit function theorem. They can be traced back to derivatives of $\mnsn,\,\mesn$ and $\mejn$. The functions $\mesn$ and $\mnsn+\mejn$ are strictly monotonically increasing in $u$ and $z$ for fixed $z$ and $u$, respectively. Thus we can solve Eqs.~\eqref{eq usimpl} and \eqref{eq ujimpl} locally to $u$, where we obtain $u^{(S)}$ and $u^{(J)}$. These equations can be derived by $z$ using
\begin{align}
 &\frac{\partial u^{(S)}}{\partial z}(\TLS,z_0)=-\left.\frac{\frac{\partial}{\partial z}\mesn(z,u)}{\frac{\partial}{\partial u}\mesn(z,u)}\right|_{u=u^{(S)}(\TLS,z_0)}\\
 \text{and}\hspace{1cm}&\frac{\partial u^{(J)}}{\partial z}(\TLJ,z_0)=-\left.\frac{\frac{\partial}{\partial z}\left(\mnsn(z,u)+\mejn(z,u)\right)}{\frac{\partial}{\partial u}\left(\mnsn(z,u)+\mejn(z,u)\right)}\right|_{u=u^{(J)}(\TLJ,z_0)}\;.
\end{align}

\subsection{Absence of subordination}
Now we consider the same model as discussed before, but without taking subordination into account. That means a loss is evenly distributed among the creditors. This model is closely related to that in Ref.~\cite{Sicking}. Here we have $B\geq 2$ creditors with the face value $\Fb,\ b=1,\dotsc,B,\;k=1,\dotsc,K$ of obligor $k$ within creditor $b$ and the normalized loss according to obligor $k$
\begin{align}
L^{(b)}_k =\begin{cases}
\left(1-\frac{V_k(T)}{F_k}\right)\Theta\left(F_k-V_k(T)\right) &\text{if } \Fb>0\\
0 & \text{else}\;,
\end{cases}
\label{eq noSubLoss}
\end{align}
with the total face value $F_k=\sum_{b=1}^{B}{\Fb}$ of obligor $k$ and asset value $V_k(t)$. In Eq.~\eqref{eq noSubLoss} for $\Fb>0$  the losses do not have any dependence on the obligors. In case of default the creditors are not distinguished and suffer the same normalized loss. Hence we write $L_k$ instead of $L^{(b)}_k$. Again we define the normalized portfolio losses $\Lb$ and the fractional face values $\fb$,
\begin{align}
\Lb=\sum\limits^{K}_{k=1}{\fb L_k}\hspace{1cm}\text{and}\hspace{1cm}\fb=\frac{\Fb}{\sum\limits^{K}_{k=1}{\Fb}}\;,
\end{align}
corresponding to creditor $b$, respectively. The multivariate distribution of the total average loss is
\begin{align}
\left\langle p\right\rangle(L)=\int{d[V]\left\langle g\right\rangle(V|\Sigma)\,\delta\!\left(L-\sum^{K}_{k=1}{f_kL_k}\right)}\;,
\label{eq NoSubStart}
\end{align}
with $L=(L^{(1)},\dots,L^{(B)})^\dagger$ and $f_k=(f^{(1)}_{k},\dots,f^{(B)}_{k})^\dagger$. Adjusting our calculations in the subordinated case above and also applying a second order approximation for $\fb$ we arrive at the final result
\begin{align}
\left\langle p\right\rangle(L) &=\frac{1}{2^{N/2}\Gamma(N/2)}\int\limits^{\infty}_{0}{dz z^{N/2-1}e^{-z/2}\sqrt{\frac{N}{2\pi }}\int\limits^{\infty}_{-\infty}{du\exp\left(-\frac{N}{2}u^2\right)}}\no\\
&\quad\times \frac{1}{\sqrt{\det\left(2\pi M_2(z,u)\right)}}\exp\left(-\frac{1}{2}\left(L-M_1(z,u)\right)^\dagger M^{-1}_2(z,u)\left(L-M_1(z,u)\right)\right)\;,
\label{eq NoSubResult}
\end{align}
where
\begin{align}
M_1(z,u)&=\sum\limits^{K}_{k=1}f_km_{1,k}(z,u)\\
M_2(z,u)&=\sum\limits^{K}_{k=1}D_k\left(m_{2,k}(z,u)-m^{2}_{1,k}(z,u)\right)
\end{align}
with the dyadic matrices
\begin{align}
 D_k=f_kf^{\dagger}_{k}
 \label{eq ddyadic}
\end{align}
and with
\begin{align}
m_{j,k}(z,u)&=\int\limits^{\hat{F}_k}_{-\infty}{d\hat{V}_k\left(1-\frac{V_{k0}}{F_k}\exp\left(\sqrt{z}\hat{V}_k+\left(\mu_k-\frac{\rho^2_k}{2}\right)T\right)\right)^j}\no\\
&\quad\times \sqrt{\frac{N}{2\pi(1-c)T\rho^2_k}}\exp\left[\frac{N}{2(1-c)T\rho^2_k}\left(\hat{V}_k+\sqrt{cT}u\rho_k\right)^2\right]\\
\hat{F}_k&=\frac{1}{\sqrt{z}}\left(\ln\frac{F_k}{V_{k0}}-\left(\mu_k-\frac{\rho^2_k}{2}\right)T\right)\;.\label{eq momentsnoSub}
\end{align}
The moments $m_{j,k}(z,u)$ are the same as in Eq.~\eqref{eq moment}. For this model we only consider heterogeneous portfolios for the whole market as homogeneous portfolios would lead to singular matrices $D_k$ as defined in Eq.~\eqref{eq ddyadic} and the losses would exactly be the same for all creditors. Instead we consider cases where the volume of credit differs among the creditors or we consider cases where the portfolios are non-overlapping or may only partially overlap.

\begin{figure}[ht]
 \centering
 \includegraphics[width=0.35\textwidth]{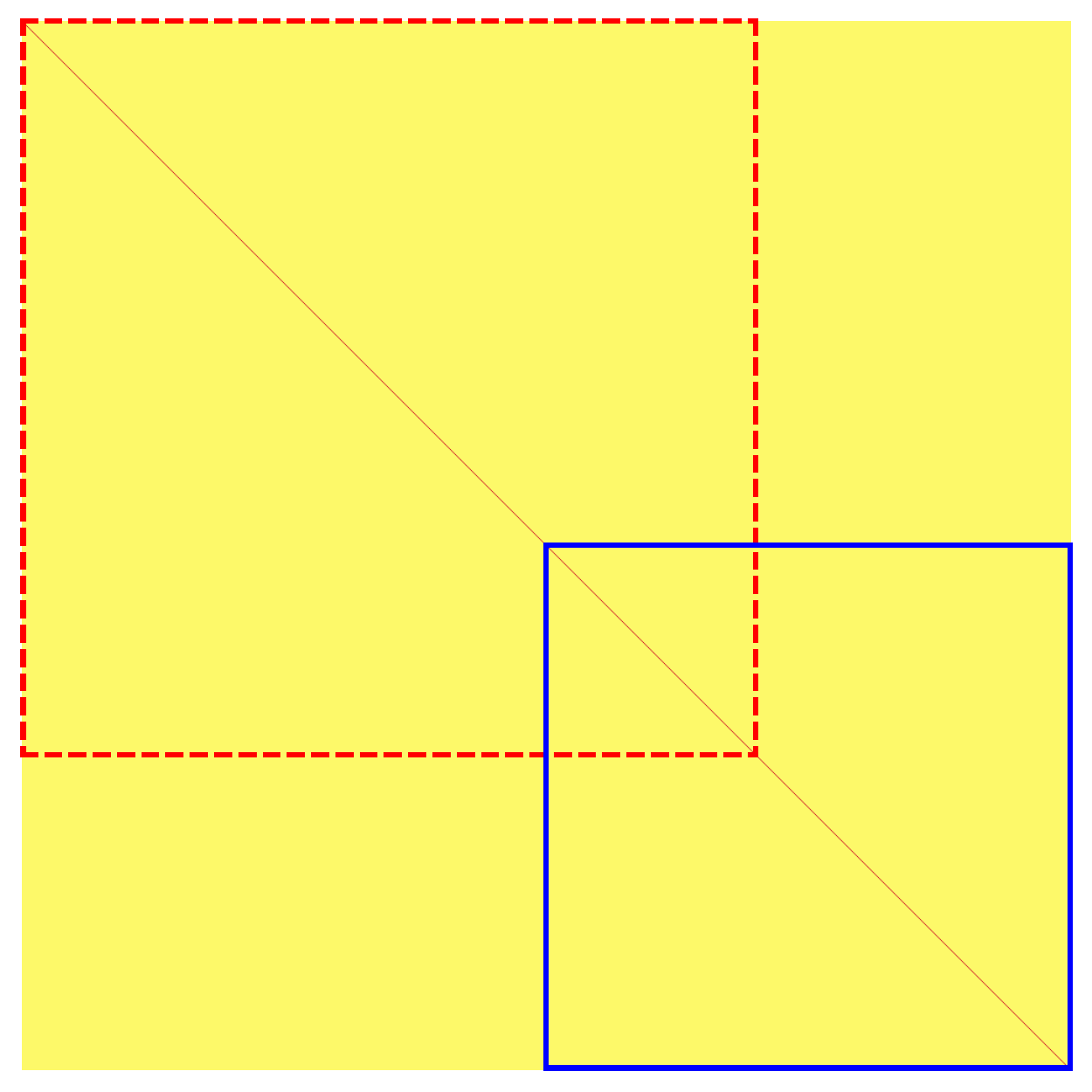}
  \caption{Setup of the generalized model illustrating the correlation matrix of a financial market with homogeneous correlations. The two rimmed squares correspond to two partially overlapping portfolios.}
 \label{fig setupNoSub}
\end{figure}

Although our results are general, we now only consider $B=2$ creditors to feasible render a visualization. We denote them as creditor one and creditor two, respectively. Moreover we address the most general setup where two credit portfolios may partially overlap, see Fig.~\ref{fig setupNoSub}.
Again we consider $K$ obligors in total. Let $R_1$ be the number of obligors with only one credit contract, say from creditor one, depicted in the upper square on the left-hand side. Let $R_{12}$ be the number of creditors that raise credits from both creditors. These creditors correspond to the overlapping area in Fig.~\ref{fig setupNoSub}. The proportions correspond to the fractions $r_1=R_1/K$ and $r_{12}=R_{12}/K$.
Creditor one deals in $R_1+R_{12}$ credits and creditor two deals in $K-R_1$ credits. This model for example also includes two disjoint portfolios, we just have to set $r_{12}=0$. The face value of the $R_{12}$ obligors consist of the sum of two face values $F_k=F^{(1)}_k+F^{(2)}_k$ that do not necessarily have the same size. For later convenience we consider homogeneous portfolios $F_k=F_0$ and we assume that the face values in the overlapping part of the portfolios are equal within a portfolio but can differ across the portfolios. That means we introduce a parameter $\gamma\in [0,1]$ with $F^{(1)}_k=\gamma F_0$ and $F^{(2)}_k=(1-\gamma)F_0$.

For a market with homogeneous parameters we find the result \eqref{eq NoSubResult} with
\begin{align}
M_1(z,u)&=m_{1,0}(z,u)\begin{bmatrix}
                            1\\
                            1
                           \end{bmatrix}\\
M_2(z,u)&=\left(m_{2,0}(z,u)-m^{2}_{1,0}(z,u)\right)\frac{1}{K}\begin{bmatrix} \alpha_1 & \alpha_{12}\\ \alpha_{12} & \alpha_{2} \end{bmatrix} \label{eq M2NoSub}
\end{align}
where
\begin{align}
\alpha_1&=\frac{r_1+\gamma^2r_{12}}{(r_1+\gamma r_{12})^2}\\
\alpha_{12}&=\frac{\gamma(1-\gamma)r_{12}}{(r_1+\gamma r_{12})(1-r_1-\gamma r_{12})}\\
\alpha_2&=\frac{1-r_1-\gamma(2-\gamma)r_{12}}{(1-r_1-\gamma r_{12})^2}\;.
\end{align}
We notice $\alpha_{12}=0$ for $\gamma=0$ or $\gamma=1$.

\subsubsection{Absence of subordination on several markets}
To treat several uncorrelated markets we perform the same calculations as in the previous section. We insert the average asset value distribution \eqref{eq assetdistseveralmarkets} into Eq.~\eqref{eq NoSubStart} with the slight difference that we have to replace the sum over $k$ by two sums over $l$ and $k$. We arrive at the final result which is up to an factor formally identical to Eq.~\eqref{eq NoSubResult}
\begin{align}
\left\langle p\right\rangle(L) &=\frac{1}{2^{N/2}\Gamma(N/2)}\int\limits^{\infty}_{0}{dz z^{N/2-1}e^{-z/2}\sqrt{\frac{N}{2\pi }}^{\beta}\int\limits{d[u]\exp\left(-\frac{N}{2}u^2\right)}}\no\\
&\quad\times \frac{1}{\sqrt{\det\left(2\pi M_2(z,u)\right)}}\exp\left(-\frac{1}{2}\left(L-M_1(z,u)\right)^\dagger M^{-1}_2(z,u)\left(L-M_1(z,u)\right)\right)
\label{eq NoSubResultSevMarkets}
\end{align}
with
\begin{align}
M_1(z,u)&=\sum^{\beta}_{l=1}\sum^{K_l}_{k=1}f_{lk}m_{1,l,k}(z,u_l)\\
M_2(z,u)&=\sum^{\beta}_{l=1}\sum^{K_l}_{k=1}D_{lk}\left(m_{2,l,k}(z,u_l)-m^{2}_{1,l,k}(z,u_l)\right)\hspace{1cm}\text{with}\hspace{1cm}D_{lk}=f_{lk}f^{\dagger}_{lk}
\end{align}
and $u=(u_1,\dotsc,u_\beta)$. Here $d[u]$ denotes the product of all differentials $du_l$. The moments $m_{1,l,k}(z,u_l)$ and $m_{2,l,k}(z,u_l)$ are the same as in Eq.~\eqref{eq momentsnoSub} including an additional index for each market $l\in\{1,\dotsc\beta\}$. In this way we are able to vary the parameters like drift and volatility across the markets. We found it useful depending on the size of $\beta$, to use polar or spherical coordinates for the evaluation of the multivariate $u$ integral.

\subsection{Absence of subordination and infinitely large portfolios}
We now consider two infinitely large portfolios, taking the limit $K\to\infty$. We point out that $r_1$ and $r_{12}$ do not scale with $K$ in the case of two infinitely large portfolios. We will consider the case of one infinitely large portfolio and one portfolio of finite size later on. Now the matrix $M_2(z,u)$ converges to a zero matrix. This implies that the exponential term and its prefactor converge to $\delta$ functions and we find the final result
\begin{align}
\lim_{K\to\infty}\left\langle p\right\rangle(L^{(1)},L^{(2)})&=\frac{1}{2^{N/2}\Gamma(N/2)}\int\limits^{\infty}_{0}{dz z^{N/2-1}e^{-z/2}\sqrt{\frac{N}{2\pi }}\int\limits^{\infty}_{-\infty}{du\exp\left(-\frac{N}{2}u^2\right)}}\no\\
&\quad\times\delta\left(L^{(1)}-m_{1,0}(z,u)\right)\delta\left(L^{(2)}-L^{(1)}\right)\;.
\label{eq NoSubLimit}
\end{align}
This result is quite remarkable. We point out first, there is no dependence on the structure of the portfolios anymore as the distribution \eqref{eq NoSubLimit} is independent of the parameters $\alpha_1,\alpha_{12}$ and $\alpha_2$. Second, in the limiting case the losses of both portfolios will always be equal to each other so that they are perfectly correlated. In other words the loss of one large creditor can be used as a forecast for the loss of another large creditor on the same market. This holds even if the creditors have disjoint portfolios and it also does not depend on the strength of the correlations across the asset values.

A different situation appears when we consider a portfolio of finite size and another infinitely large one. Due to the high asymmetry of the market shares of the portfolios, we solely examine two disjoint portfolios. Say, portfolio one is the finite one with $R_1$ companies. Then the matrix element $\alpha_1$ in Eq.~\eqref{eq M2NoSub} scales with $K$ and $\alpha_2$ converges to one. By calculating the limit $K\to\infty$ only one $\delta$ function emerges and by using property \eqref{eq deltatrick} of the $\delta$ function we find
\begin{align}
\lim_{K\to\infty}\left\langle p\right\rangle(L^{(1)},L^{(2)})&=\frac{1}{2^{N/2}\Gamma(N/2)}\int\limits^{\infty}_{0}{dz z^{N/2-1}e^{-z/2}\sqrt{\frac{N}{2\pi}}\exp\left(-\frac{N}{2}u_0^2\right)}\no\\
&\quad\times\sqrt{\frac{R_1}{2\pi(m_{2,0}(z,u_0)-m^{2}_{1,0}(z,u_0))}}\exp\left(-\frac{R_1(L^{(1)}-m_{1,0}(z,u_0))^2}{2(m_{2,0}(z,u_0)-m^{2}_{1,0}(z,u_0))}\right)\no\\
&\quad\times\frac{1}{\left|\partial m_{1,0}(z,u)/\partial u\right|_{z,u_0}}\;,
\end{align}
where $u_0(L^{(2)},z)$ is an implicit function defined by
\begin{align}
0=L^{(2)}-m_{1,0}(z,u_0)\;.
\end{align}
We note that the dependence on $L^{(2)}$ in the limit distribution is encoded in $u_0(L^{(2)},z)$. Moreover the above result is in line with the second order approximation even though one of the matrix elements does not scale with $K$.

Finally we analyze two disjoint infinitely large portfolios, where each portfolio invests on a separate market. We start from distribution \eqref{eq NoSubResultSevMarkets} and perform the limit $K\to\infty$. Again we find two $\delta$ functions and by applying Eq.~\eqref{eq deltatrick} twice we obtain
\begin{align}
\lim_{K\to\infty}\left\langle p\right\rangle(L^{(1)},L^{(2)})&=\frac{1}{2^{N/2}\Gamma(N/2)}\int\limits^{\infty}_{0}{dz z^{N/2-1}e^{-z/2}\frac{N}{2\pi}\exp\left(-\frac{N}{2}u_{10}^2\right)\exp\left(-\frac{N}{2}u_{20}^2\right)}\no\\
&\quad\times\frac{1}{\left|\partial m_{1,1,0}(z,u)/\partial u\right|_{z,u_{10}}}\frac{1}{\left|\partial m_{1,2,0}(z,u)/\partial u\right|_{z,u_{20}}}\;,
\end{align}
where 
\begin{align}
0=L^{(1)}-m_{1,1,0}(z,u_{10})\hspace{1cm}\text{and}\hspace{1cm}0=L^{(2)}-m_{1,2,0}(z,u_{20})
\end{align}
define the implicit functions $u_{10}(L^{(1)},z)$ and $u_{20}(L^{(2)},z)$.

\section{Calibration of our model and visualization of the results}\label{Sec results}
We always employ the approximation \eqref{eq avcorrm} to the mean correlation matrix, which yields as already emphasized, very good fits to empirical data due to the very nature of the ensemble average. Furthermore we restrict our analysis to homogeneous portfolios.

\subsection{One portfolio, two markets}
There are four parameters in our model, the average drift $\mu$, the average volatility $\rho$, the average correlation level $c$ and the parameter $N$ which controls the strength of the fluctuations around the mean correlation level. In Refs.~\cite{schmitt2014EPL,schmitt2015CR,schmitt2013EPL} the values of the parameters are directly estimated from empirical data. The parameter $N$ is determined by a fit of distribution \eqref{eq dist rets} to the data. These consists of 307 stocks that are taken from the S\&P500 index traded in the period from 1992 to 2012 \cite{yahoo}. We find the following empirical results $\mu=0{.}17\,\text{year}^{-1}$, $\rho=0{.}35\,\text{year}^{-1/2}$, $N=6$, $c=0{.}28$, $T=1\,\text{year}$. 

\begin{figure}[ht]
 \centering
 \includegraphics[width=0.7\textwidth]{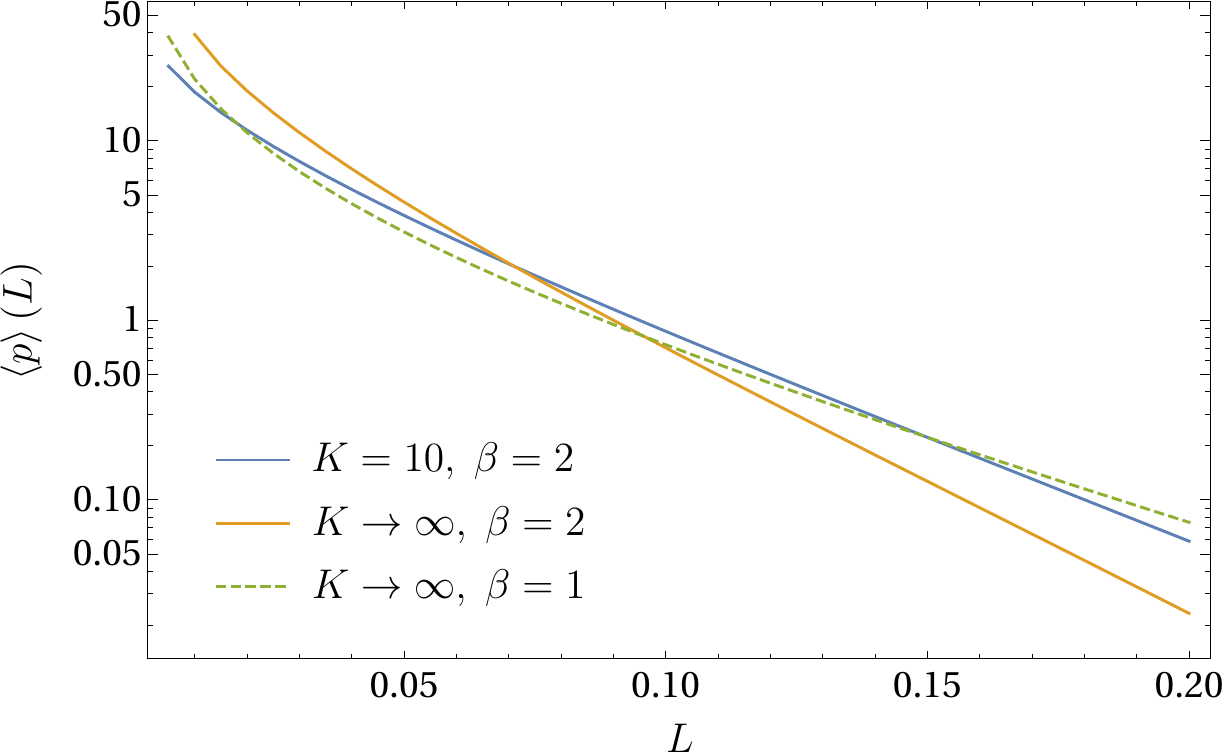}
  \caption{Average loss distribution on a logarithmic scale for different number of markets and market size. The markets are homogeneous and we choose a face value of $F_0=75$ and the initial asset value $V_0=100$.}
 \label{fig 2MUncorr}
\end{figure}

We study the impact of investing into two uncorrelated markets. We thus assume two identical uncorrelated markets with the same average correlation level. Furthermore we assume the empirical parameters to be the same for both markets. The results are shown in Fig.~\ref{fig 2MUncorr}.
For a comparison we also show the limiting distribution for only one market $\beta=1$ with the same parameters as in the case of two markets. This distribution is the univariate version of the distribution \eqref{eq NoSubLimit} see also Ref.~\cite{schmitt2015CR}. As expected we see that the diversification, i.e. the separation of the correlation matrix into two blocks leads to a reduction of large portfolio losses. Hence, reducing the risk of large losses can be achieved more effectively by splitting the portfolio onto different uncorrelated markets than by solely increasing the number of credit contracts on one single market. Obviously this is due to the on average zero correlations in the off-diagonal blocks. A further reduction of the risk can only be achieved by either splitting the portfolio in more than two markets or investing into markets where the average correlation level is low with little fluctuations. Nevertheless by increasing the number of uncorrelated markets $\beta$ we obtain for $\beta\to\infty$ the same scenario as in the case of one market with average correlation zero. Here the diversification effect is limited to the strength of the fluctuations $N$, where the tail of the loss distribution would only vanish for large $N$.

\subsection{Absence of subordination and disjoint portfolios of equal size}\label{subsec: NoSubEq}
\begin{figure}[ht]
 \centering
 \includegraphics[width=0.7\textwidth]{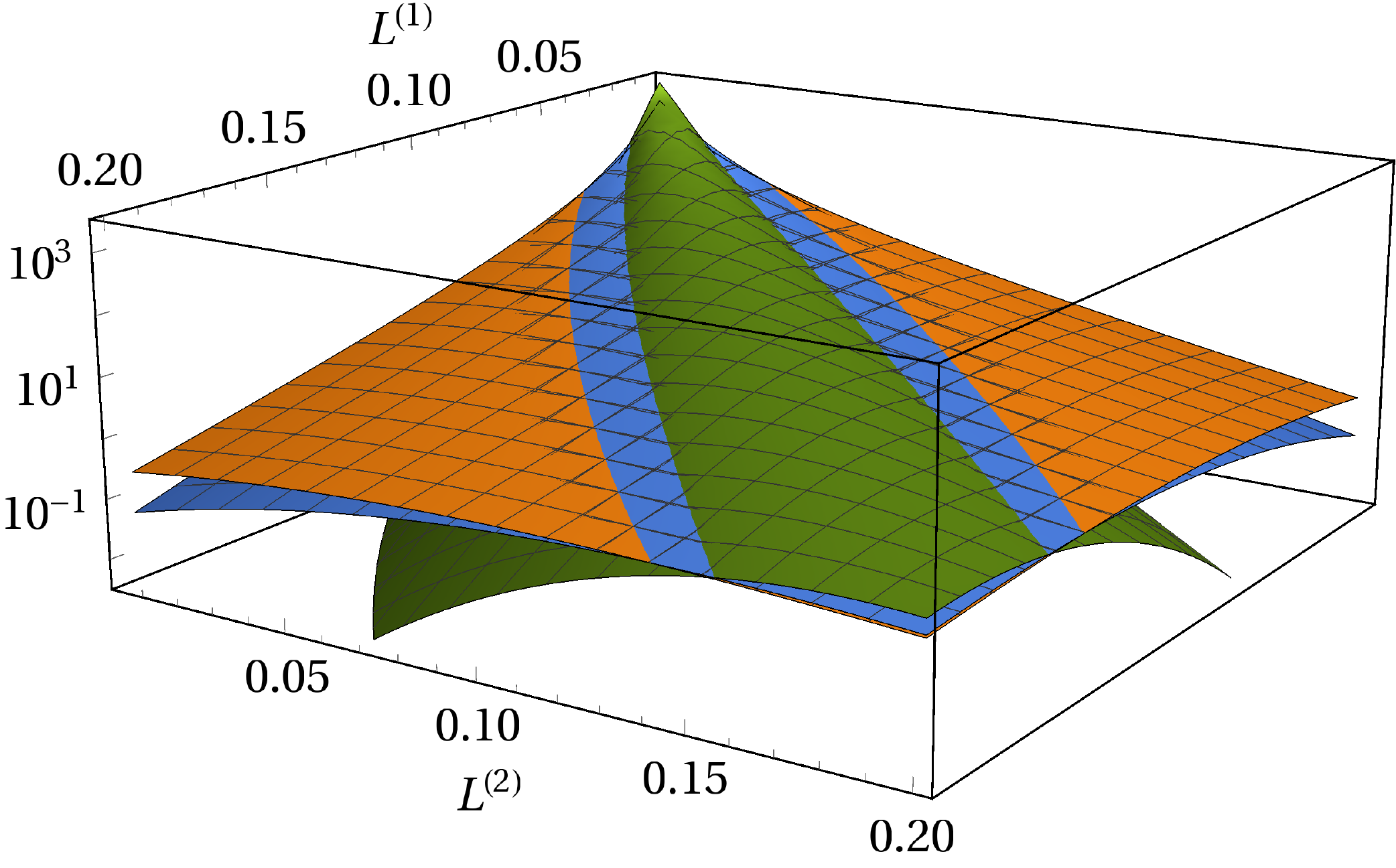}
  \caption{Average loss distribution for two disjoint portfolios of same sizes on a logarithmic scale. We show different market sizes, $K=10$ orange, $K=20$ blue and $K=100$ green. The parameters are $\mu=0{.}17\,\text{year}^{-1}$, $\rho=0{.}35\,\text{year}^{-1/2}$, $N=6$, $c=0{.}28$ and a maturity time of $T=1\,\text{year}$.}
 \label{fig NoSubEq}
\end{figure}

\begin{figure}[ht]
 \centering
 \includegraphics[width=0.7\textwidth]{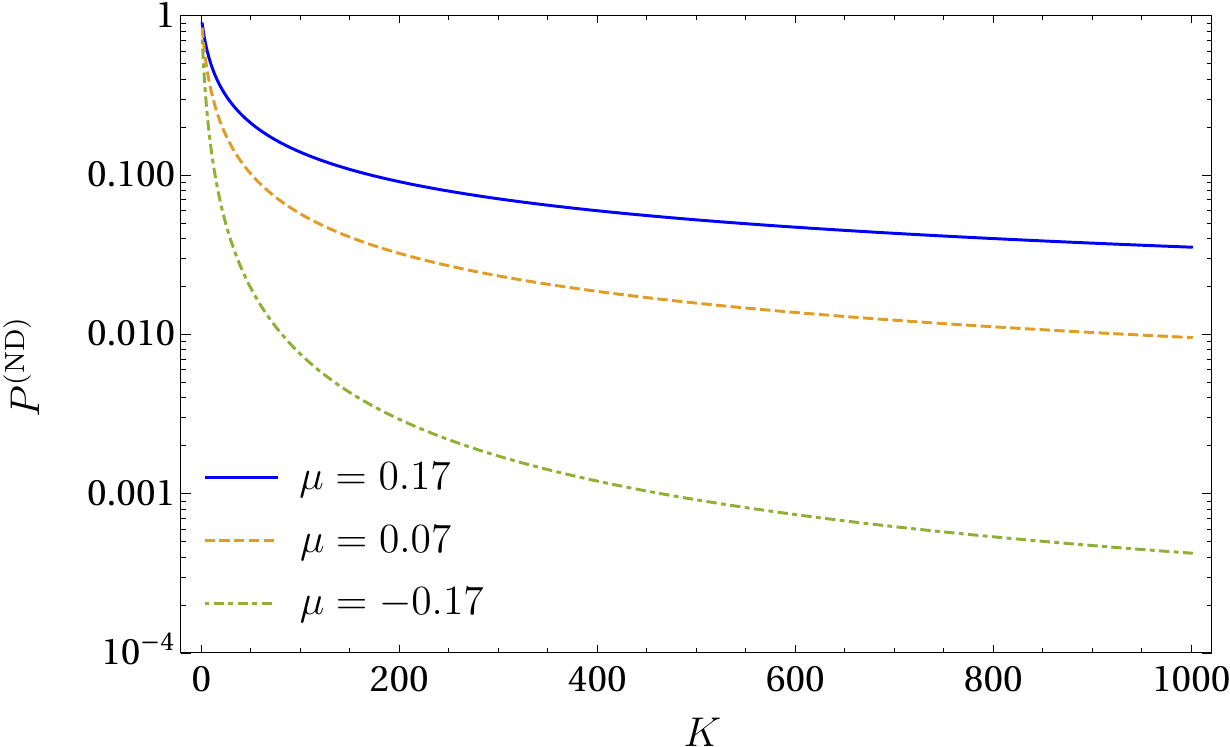}
  \caption{Probability of zero portfolio loss depending on the portfolio size $K$ for different drift parameters $\mu$ on a logarithmic scale.}
 \label{fig DefProb}
\end{figure}

We begin with varying the number of companies $K$ and study the impact on the multivariate loss distribution as well as on the default correlation and the default probabilities. Fig.~\ref{fig NoSubEq} shows the average loss distribution \eqref{eq NoSubResult} with homogeneous correlation matrix and homogeneous parameters for two disjoint portfolios of equal size, for different numbers of companies $K=10,20,100$ and empirical values for the parameters.
We choose the face value $F_0=75$ and the initial asset value $V_0=100$. The distribution is symmetric. It converges to the limiting distribution \eqref{eq NoSubLimit} as $K$ increases. We thus infer a high correlation of the portfolio losses even for a small number of obligors. The striking peak around the origin $L^{(1)}=L^{(2)}=0$ corresponds to those events that lead to little portfolio loss.  This peak arises because of the large drift. Due to the positive drift the overall number of companies which do not default is larger than the number of companies that default at maturity. Still, this peak does not represent the $\delta$ peak at the origin which stands for the probability of total survival of all companies. This becomes clear when we calculate the survival probability for all companies. This probability does not depend on whether we have subordinated debt or not and it also does not depend on the composition of our portfolios, see Eq.~\eqref{eq survivalprobab}. The effect of different drift parameters $\mu$ is shown in Fig.~\ref{fig DefProb}.
For every value of $\mu$ the probability of having zero total portfolio loss decreases with increasing number of companies $K$. Hence the weight of the $\delta$ peak on the portfolio loss distribution at $L^{(1)}=L^{(2)}=0$ becomes smaller. This is quite intuitive, the larger $K$ the more likely is the default of at least one company.

\begin{figure}[ht]
 \centering
 \includegraphics[width=0.7\textwidth]{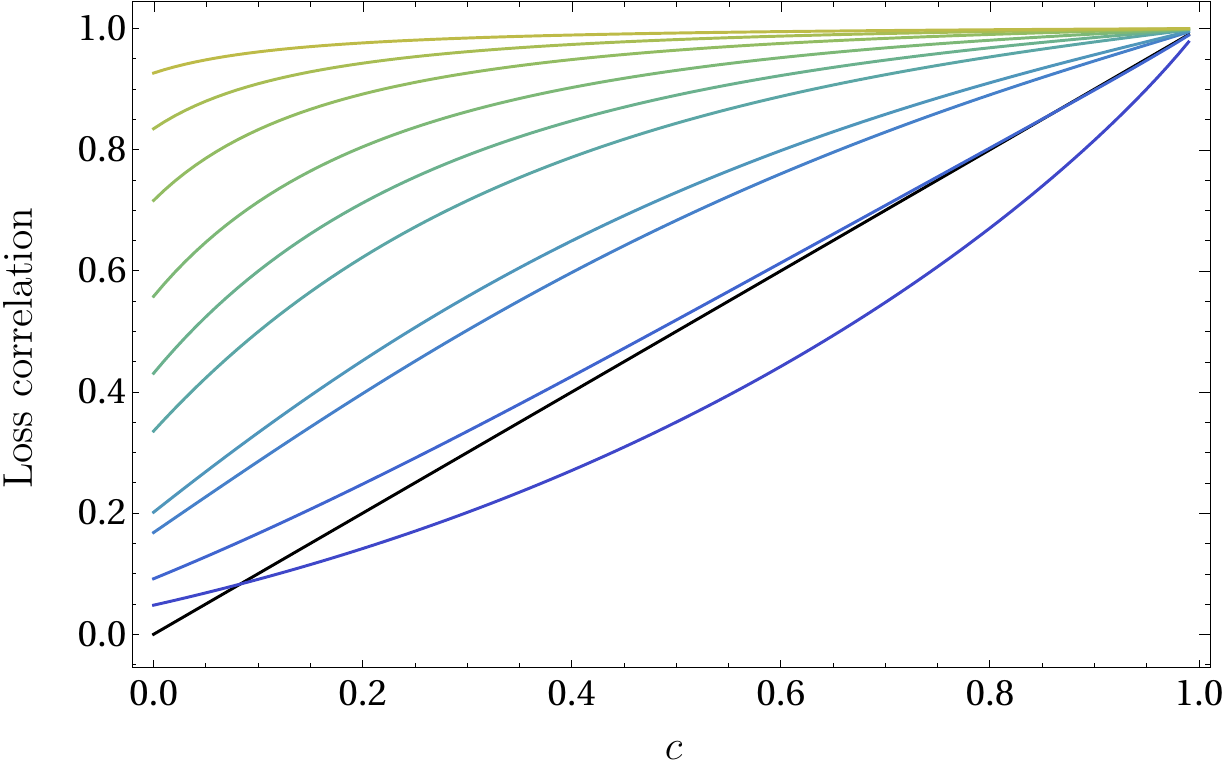}
  \caption{Portfolio loss correlation on a linear scale depending on the asset correlation $c$. Both portfolios are homogeneous and have the same size. The market size $K$ ranges from 2 (blue) over 4,8,10,20,30,50,100,200 to 500 (green). The bisecting line is shown in black.}
 \label{fig CorrEq}
\end{figure}

When looking at the portfolio loss correlations we find large values for little or even zero asset correlation. For a market size of $K=100$, i.e. each portfolio is of size 50, we obtain for an average asset correlation of $c=0$ a correlation of the portfolio losses of $0{.}71$. This high loss correlation is based on the fact that the asset correlations fluctuate around the mean asset correlation of zero. Due to this fluctuation we have individual positive and negative correlations. The negative correlations only have a limited effect because the asymmetry of credit risk projects all non defaulting events onto zero while only defaulting events contribute to the loss distribution. Hence, the positive asset correlations dominate the negative ones causing a high portfolio loss correlation. The results is shown in Fig.~\ref{fig CorrEq}.
They are in accordance with the simulation results in Ref.~\cite{Sicking}. The portfolio loss correlation is a monotonic function of the asset correlation $c$ and for a fixed asset correlation we find with increasing $K$ an increasing portfolio loss correlation. Depending on the number of companies, the portfolio loss correlation is a convex function (namely, $K=2,4$) or a concave function ($K\geq 8$). However, we emphasize that these results are subject to the second order approximation which yields better results the larger $K$. Large numbers of $K$ lead to very high loss correlations. This confirms that even without average asset correlation $c=0$ the loss of one large portfolio serves as a forecast for another large portfolio.

\subsection{Absence of subordination and disjoint portfolios of various size}
Looking at portfolios of various sizes yields much improved understanding of whether diversification works or not. To analyze this we consider portfolio one with fixed size $R_1=10$ and we consider the overall size of the market $K=30,110$ and the limit $K\to\infty$. In this scenario the market share of portfolio one will steadily decrease and converge to zero in the limiting case. Hereby, we are able to compare somewhat smaller portfolios with very large ones.

\begin{figure}[ht]
 \centering
 \includegraphics[width=0.7\textwidth]{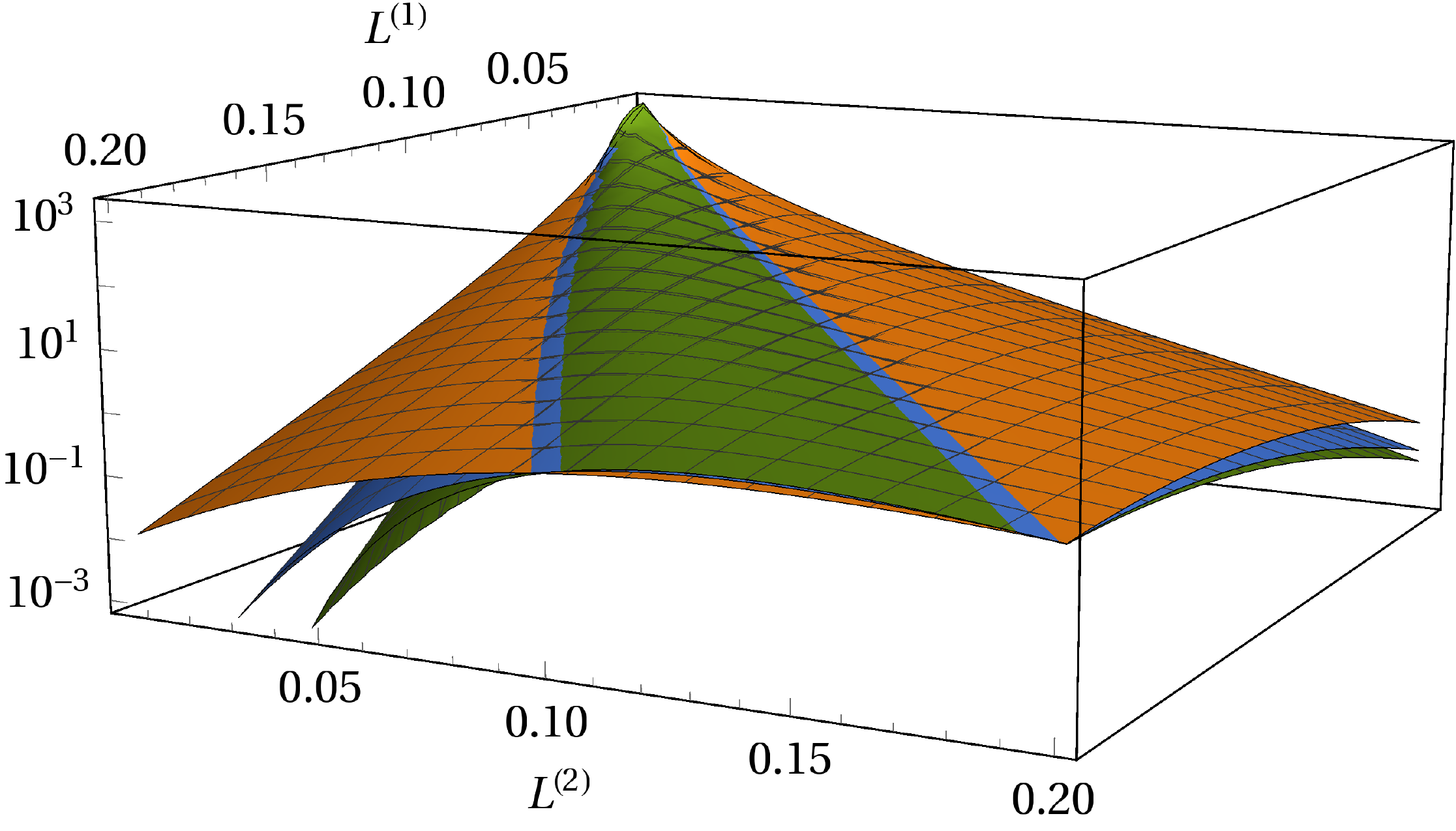}
  \caption{Average loss distribution for portfolios of different sizes on a logarithmic scale. Portfolio one is of fixed size $R_1=10$ and the market size is $K=30$ (orange), $K=110$ (blue) and $K\to\infty$ (green). The parameters are $\mu=0{.}17\,\text{year}^{-1}$, $\rho=0{.}35\,\text{year}^{-1/2}$, $N=6$, $c=0{.}28$ and a maturity time of $T=1\,\text{year}$.}
 \label{fig pdfDiffSz}
\end{figure}

\begin{figure}[ht]
 \centering
 \includegraphics[width=0.7\textwidth]{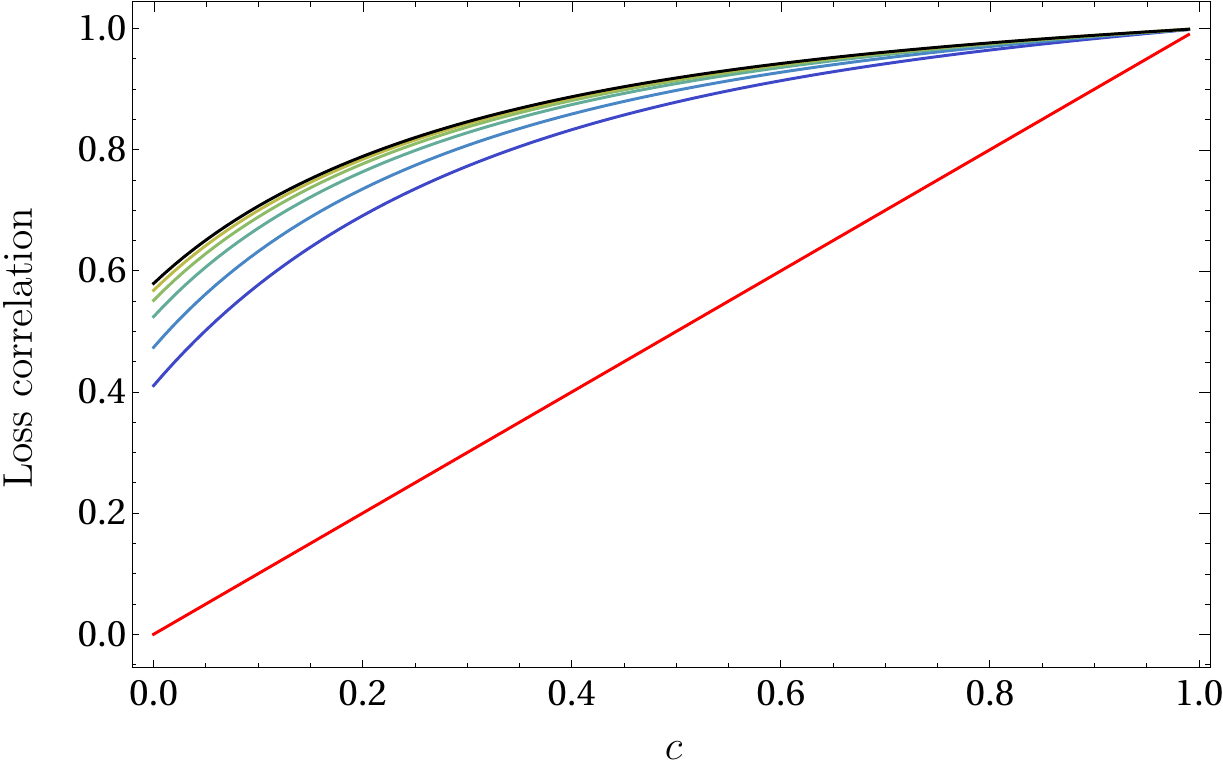}
  \caption{Portfolio loss correlation as a function of asset correlation $c$ on a linear scale. Portfolio one is of fixed size $R_1=10$ and the market size $K$ ranges from 30 (blue) over 50, 100, 200 to 500 (green). The limiting curve $K\to\infty$ is shown in black and the bisecting line is shown in red.}
 \label{fig CorrVar}
\end{figure}

For our calculations we use the same empirical parameters as in section \ref{subsec: NoSubEq}. The effect of different market size $K$ on the loss distribution $\left\langle p\right\rangle(L^{(1)},L^{(2)})$ is shown in Fig.~\ref{fig pdfDiffSz}.
There are regions where we have heavy-tailed behaviour of the distributions but also others where the distributions decay very fast. In this latter regions that always fulfill the condition $L_1>L_2$ the loss distribution decays considerably faster with increasing market size $K$. Hence we find large deviations between the distributions of different market sizes. These deviations only play a minor role because they emerge at a significant low order of the loss distribution. In general, for increasing market size $K$ the second portfolio describes the market in a better manner. Hence it is very unlikely for the first portfolio to suffer a big loss in times when the second portfolio of large size exhibits little loss. This explains the fast decay of the loss distribution in the $L_1>L_2$ corner. However the most important fact is that along the diagonal $L_1=L_2$ and in the upper corner $L_1<L_2$, significant deviations between the loss distributions for different market sizes do not occur. Here, we also observe heavy-tails of the loss distribution. Especially when we consider the diagonal we find no deviations and thus no diversification at all. This means that increasing the size of portfolio 2 while keeping the size of portfolio 1 constant does not yield a decrease of concurrent large portfolio losses of equal size. Interestingly, it is more likely to find an event in the upper off-diagonal corner with $L_1<L_2$ than in the lower corner. This can be explained by the fluctuations around the mean correlation level of $c=0{.}28$ and the positive drift $\mu=0{.}17\,\text{year}^{-1}$. The fluctuations ensure that there is a probability for the assets of portfolio one to be adversely correlated to the assets of portfolio two. Accordingly there is a significant probability that the small portfolio one suffers no or little default while the second portfolio suffers a major one. This probability decreases when we enlarge the size of portfolio one while keeping the size of portfolio two fixed and still larger than the size of portfolio one. Due to the asymmetry of the portfolio loss distributions regarding the diagonal we find lower loss correlation for the same market size than in the case of two equal sized portfolios, see Fig.~\ref{fig CorrVar}.
In contrast to two portfolios of equal size there is a limit correlation of the portfolio losses depending on $c$ in the limit $K\to\infty$. One clearly sees that the limiting curve is reached very quickly for increasing market size. This is due to the fixed size of portfolio one. Increasing its size and the market size would raise the portfolio loss correlation.

\subsection{Subordinated debt}
\begin{figure}[ht]
 \centering
 \includegraphics[width=0.7\textwidth]{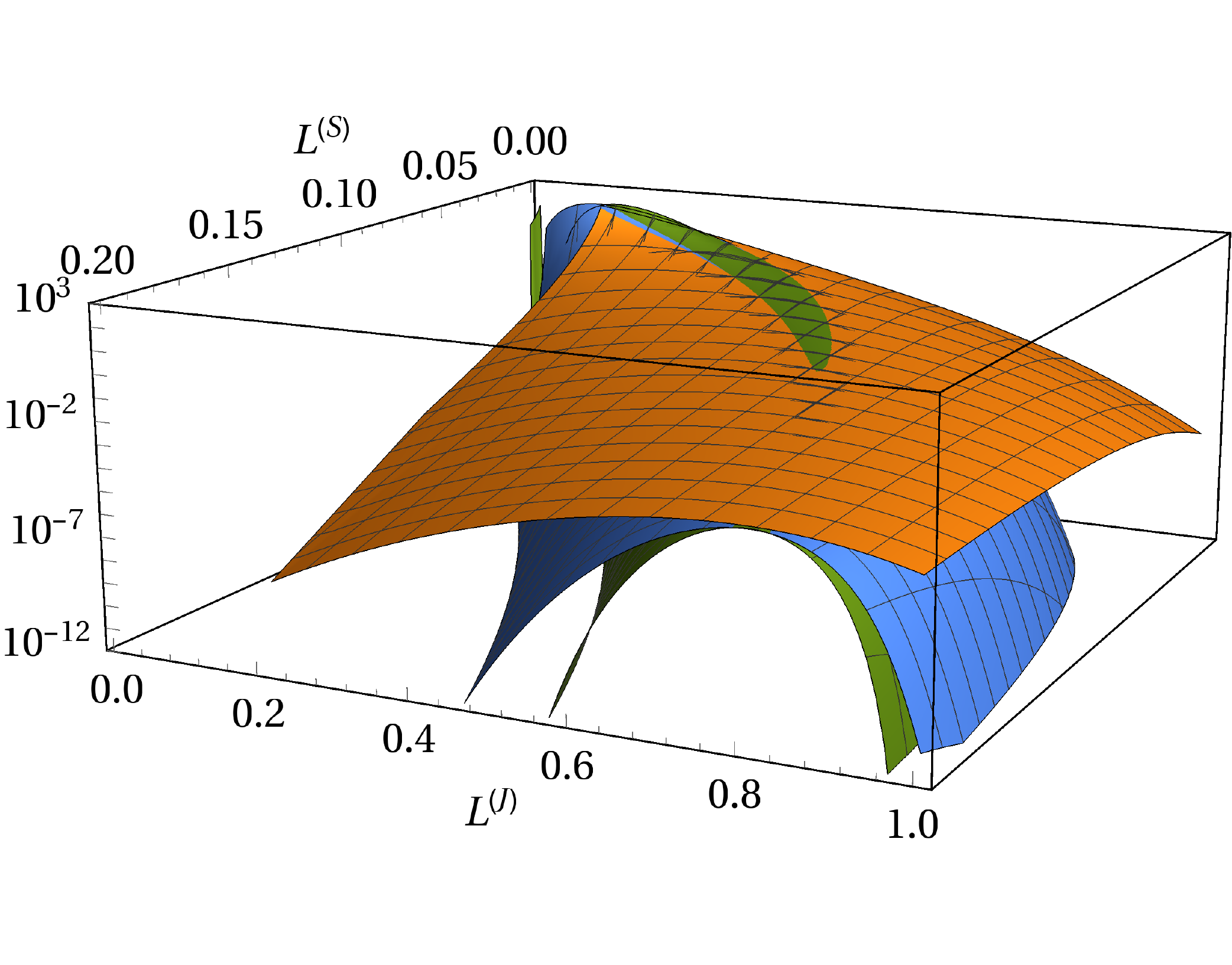}
  \caption{Average portfolio loss distribution of a subordinated debt structure on a logarithmic scale. Both portfolios operate on the entire market. We show different market sizes $K=10$ orange, $K=200$ blue and $K\to\infty$ green.}
 \label{fig SubEqual}
\end{figure}

The subordinated debt structure brings a high degree of asymmetry into effect, see Fig.~\ref{fig SubEqual}. We show the joint probability density of two equal-sized portfolios with face values $\FS=37$ and $\FJ=38$. Both, senior and junior subordinated creditor, operate on the entire market. 
The loss of the junior subordinated creditor is always larger or equal than the loss of the senior creditor. We thus have an cutoff along the diagonal line $\TLS=\TLJ$. Besides the near region of a curved line, which we define as the back of the distribution, the number of obligors $K$ influences the joint probabilities drastically. Along the back of the distribution there is almost no deviation between the surfaces of the joint probability densities. Independent of $K$, the back of the distribution shows heavy tails. Importantly the curvature reaches for high losses of the junior subordinated creditor evermore to higher losses of the senior creditor. This is an important consequence in times of crisis. When the loss of the junior subordinated creditor becomes extremely large it is most likely that also the senior creditor suffers a significant loss.

\begin{figure}[ht]
 \centering
 \includegraphics[width=0.7\textwidth]{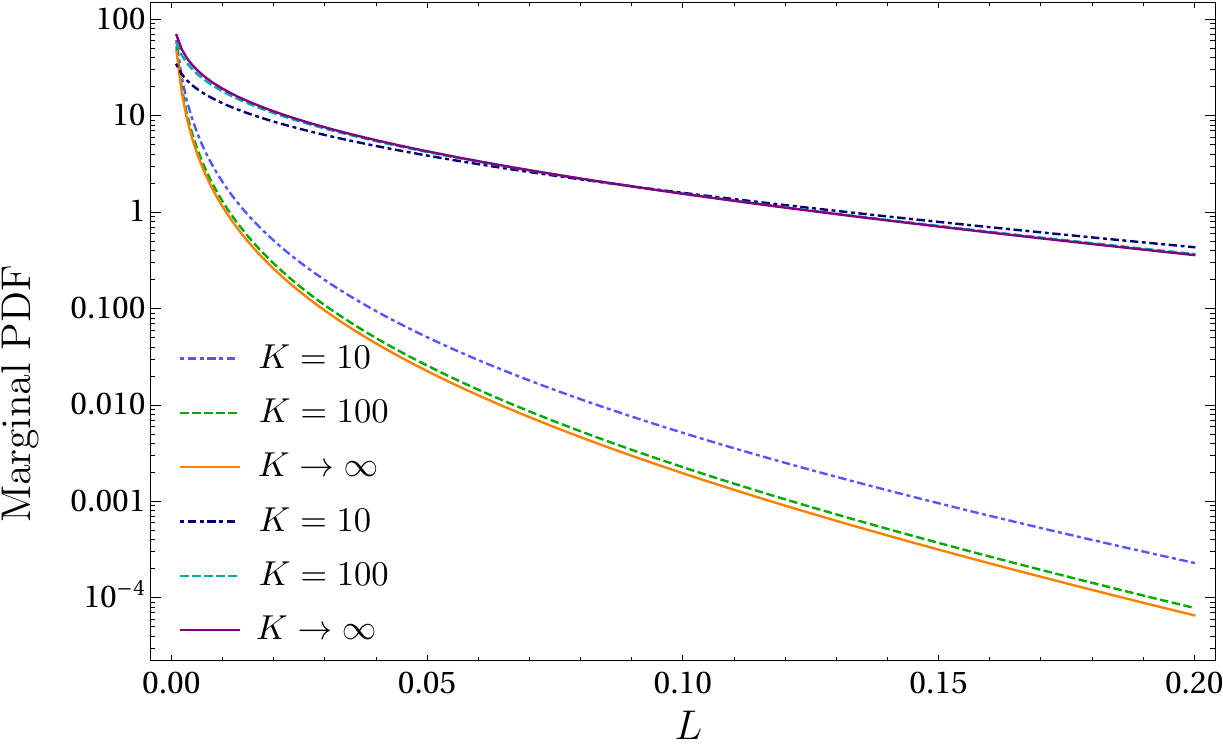}
  \caption{Marginal distributions of senior and junior subordinated creditor on a logarithmic scale. The upper three lines belong to the junior subordinated creditor and the lower three lines to the senior creditor.}
 \label{fig SubMarg}
\end{figure}

This explains why strong diversification effects do not exist, when we consider the marginal distributions of each creditor, see Fig.~\ref{fig SubMarg}.
The upper three curves belong to the marginal distributions of the junior subordinated creditor and the lower three curves belong to the senior creditor. All distributions show heavy tails and the gap between the senior and junior subordinated creditor enlarges with increasing loss $L$. The size of this gap becomes smaller when the ratio $\FS/\FJ$ becomes larger. 

\section{Conclusions}\label{Sec conclusion}
Within the Merton model we calculated an multivariate joint average portfolio loss distribution, taking fluctuating asset correlations into account. We used a random matrix model that is, most advantageously, analytically tractable and also empirically a good match of stock market data. The multivariate average asset value distribution depends on two parameters only, the effective average asset value correlation and the strength of the fluctuations around this average. 

We showed that diversification is achieved much more efficiently by splitting a credit portfolio onto different markets that are on average uncorrelated, than by solely increasing the number of credit contracts on one single market.

For two non-overlapping portfolios of equal size we found a symmetric portfolio loss distribution. Studying the portfolio loss correlations we showed that significant correlations emerge not only for large portfolios containing thousands of credit contracts but also, in accordance with a second order approximation, for small portfolios containing only a few credit contracts. Two non-overlapping portfolios of infinite size have a loss correlation of one and will always suffer the same relative loss. 

When we analyzed two non-overlapping portfolios of different size we found the loss correlations to be limited. Nevertheless, the distributions show heavy tails which make large concurrent portfolio losses likely.

Furthermore, we included subordinated debt, related to CDO tranches. At maturity time the senior creditor is paid out first and the junior subordinated creditor only if the senior creditor regained the full promised payment. Here, we analytically derived that in case of crisis, i.e. when a large loss of the junior subordinated creditor is highly likely, a large loss of the senior creditor is also very likely. Thus, the concept of subordination does not work as intended in times of crisis. In addition, the marginal distributions show that increasing the size of both portfolios fails to reduce the tail risk significantly. 

\section{Acknowledgments}
We thank Martin T.~Hibbeln, R{\"u}diger Kiesel and Sebastian M.~Krause for fruitful discussions.

\pagebreak

\appendix

\section{Moments}\label{app: moments}
We define
\begin{align}
 \tau^{\iota,\lambda}_{j,k}(z,u)&=\int\limits^{\hat{F}^{(\lambda)}_k}_{-\infty}{d\hat{V}_k\left(c^{(\iota)}-\frac{V_{k0}}{\hat{F}^{(\lambda)}_k}\exp\left(\sqrt{z}\hat{V}_k+\left(\mu_k-\frac{\rho^2_k}{2}\right)T\right)\right)^j}\no\\
&\quad\times \sqrt{\frac{N}{2\pi(1-c)T\rho^2_k}}\exp\left[\frac{N}{2(1-c)T\rho^2_k}\left(\hat{V}_k+\sqrt{cT}u\rho_k\right)^2\right]\;,
\end{align}
where $\iota=S,J$ and $\lambda=S,J$, as well as $c^{(S)}=1$ and $c^{(J)}=\frac{F_k}{F^{(J)}_{k}}$. Hence, we can write the moments
\begin{align}
 m^{(S)}_{j,k}(z,u)&=\tau^{S,S}_{j,k}(z,u)\\
 m^{(J)}_{j,k}(z,u)&=\tau^{J,J}_{j,k}(z,u)-\tau^{J,S}_{j,k}(z,u)\;.
\end{align}
With the following definition
\begin{align}
 \Phi(x)=\frac{1}{2}+\frac{1}{2}\erf\left(\frac{x}{\sqrt{2}}\right)
\end{align}
and the error function
\begin{align}
 \erf(x)=\frac{2}{\sqrt{\pi}}\int\limits^{x}_{0}{dxe^{-x^2}}
\end{align}
we can express the quantities $\tau^{\iota,\lambda}_{j,k}(z,u)$ for $j=0,1,2$
\begin{align}
 \tau^{\iota,\lambda}_{0,k}(z,u)&=\sqrt{\frac{N}{2\pi(1-c)T\rho^2_k}}\int\limits^{\hat{F}^{(\lambda)}_k}_{-\infty}{d\hat{V}_k\exp\left[\frac{N}{2(1-c)T\rho^2_k}\left(\hat{V}_k+\sqrt{cT}u\rho_k\right)^2\right]}\no\\
 &=\Phi\left(\sqrt{\frac{N}{(1-c)T\rho^2_k}}\left(\hat{F}^{(\lambda)}_k+\sqrt{cT}u\rho_k\right)\right)\\ 
\tau^{\iota,\lambda}_{1,k}(z,u)&=c^{(\iota)}\tau^{\iota,\lambda}_{0,k}(z,u)-\frac{V_{k0}}{F^{(\iota)}_k}\exp\left[\frac{z(1-c)T\rho^2_k}{2N}-\sqrt{zcT}u\rho_k+\left(\mu_k-\frac{\rho^2_k}{2}\right)T\right]\no\\
&\quad\times\Phi\left(\sqrt{\frac{N}{(1-c)T\rho^2_k}}\left(\hat{F}^{(\lambda)}_k+\sqrt{cT}u\rho_k\right)-\sqrt{\frac{z(1-c)T\rho^2_k}{N}}\right)\\
\tau^{\iota,\lambda}_{2,k}(z,u)&=-{c^{(\iota)}}^2\tau^{\iota,\lambda}_{0,k}(z,u)+2c^{(\iota)}\tau^{\iota,\lambda}_{1,k}(z,u)\no\\
&\quad+\frac{V^2_{k0}}{F^{{(\iota)}^2}_k}\exp\left[\frac{2z(1-c)T\rho^2_k}{N}-2\sqrt{zcT}u\rho_k+2\left(\mu_k-\frac{\rho^2_k}{2}\right)T\right]\no\\
&\quad\times\Phi\left(\sqrt{\frac{N}{(1-c)T\rho^2_k}}\left(\hat{F}^{(\lambda)}_k+\sqrt{cT}u\rho_k\right)-2\sqrt{\frac{z(1-c)T\rho^2_k}{N}}\right)\;.
\end{align}

\end{document}